\documentclass[preprintnumbers,amsmath,amssymb,floatfix,10pt,prd,onecolumn,
superscriptaddress,nofootinbib]{revtex4}

\usepackage[parfill]{parskip}    
\usepackage{graphicx}
\usepackage{amssymb}
\usepackage{epstopdf}
\usepackage{color}
\usepackage{float}
\usepackage{amsmath}
\usepackage{appendix}
\usepackage{hyperref}
\usepackage{orcidlink}

\usepackage{graphicx}
\usepackage{pdfpages}

\begin{document}

\title{Minimally deformed wormholes inspired by noncommutative geometry    }

\author{Francisco Tello-Ortiz \orcidlink{0000-0002-7104-5746}}
\email{francisco.tello@ua.cl}
\affiliation{Departamento de F\'isica, Facultad de Ciencias Básicas, Universidad de Antofagasta, Casilla 170, Antofagasta, Chile.}

\author{B. Mishra \orcidlink{0000-0001-5527-3565}}
\email{bivu@hyderabad.bits-pilani.ac.in}
\affiliation{Department of Mathematics, Birla Institute of Technology and Science-Pilani, Hyderabad Campus, Hyderabad 500078, India.}
\author{A. Alvarez}
\email{aalvarezu88@gmail.com}
\affiliation{Unidad de Equipamiento Cient\'ifico (Maini), Universidad Cat\'olica del Norte, 
Av. Angamos 0610, Antofagasta, Chile.}

\author{Ksh. Newton Singh}
\email{ntnphy@gmail.com}
\affiliation{Department of Physics, National Defence Academy, Khadakwasla, Pune-411023, India.}

\begin{abstract}
We introduce the\ldots
\end{abstract}
\begin{abstract}
{In this article, new wormhole solutions in the framework of General Relativity are presented. Taking advantage of gravitational decoupling by means of minimal geometric deformation approach and, the so--called noncommutative geometry Gaussian and Lorentzian density profiles, the seminal Morris--Thorne space--time is minimally deformed providing new asymptotically wormhole solutions. Constraining the signature of some parameters, the dimensionless constant $\alpha$ is bounded using the flare--out and energy conditions. In both cases, this results in an energy--momentum tensor that violates energy conditions, thus the space--time is threading by exotic matter. However, it is possible to obtain a positive defined density at the wormhole throat and its neighborhood. To further support the study a thoroughly graphical analysis has been performed.}
\end{abstract}
\maketitle

\section{Introduction}\label{sec1}

{Wormholes are exotic topological solutions of the Einstein field equations. These topological entities connect two points of the same or different universes, via a throat (tunnel) \cite{Morris1988,Morris1988A}. To allow matter to pass (traversability) from one end of the wormhole to the other, it is necessary to incorporate the idea of ``exotic matter" into the solution \cite{Morris1988,Morris1988A,Visser:1995cc}. Since they violate the so--called energy conditions \cite{Morris1988}, in principle wormhole space--times are not physically feasible solutions in the framework of General Relativity (GR hereinafter). Nevertheless, the mentioned condition cannot be surpassed or avoided because if the matter distribution is corresponding to a mater field respecting energy conditions, then the wormhole is not traversable. This is so because, the minimal geometric feature for this kind of solution being a traversable wormhole, is the satisfaction of the so--called flare--out condition. Therefore, within the purview of GR, wormhole solutions cannot exist if both the energy conditions and the flare--out condition are simultaneously satisfied \cite{Hochberg:1997wp,Visser:1997yn,Hochberg:1998ii}. It is worth mentioning that in general this fact comes directly from the analysis of the Einstein field equations \cite{Morris1988,Morris1988A,Visser:1995cc}.      }

{Regarding the previous point, a lot of articles have been published in the last three decades in order to understand how these solutions work in the context of GR and its extensions/modifications. For example inflating Lorentzian wormholes were analyzed in \cite{Roman:1992xj}, whereas evolving Lorentzian wormholes and traversable wormholes with traceless matter were studied in \cite{Kar:1994tz,Kar:1995ss} and \cite{Kar:1995vm}, respectively. Furthermore, the seminal Morris--Thorne solution \cite{Morris1988,Morris1988A} was analyzing in the presence of cosmological constant \cite{Lemos:2003jb}. Also, using non conventional matter distributions such as a phantom field \cite{Lobo:2005us,Lobo:2005yv} and Chaplygin gas \cite{Lobo:2005vc} wormhole solutions and their stability were studied. Moreover, incorporation of the so--called conformal motion symmetry, some interesting wormhole solutions have been found in \cite{Boehmer:2007md,Boehmer:2007rm,Rahaman2013,Kuhfittig2016,Kuhfittig2017}.}

{Besides the common procedure \emph{a la} Morris--Thorne to build up wormhole structures, in \cite{Visser:1989kg,Poisson:1995sv} the so--called cut and paste method to generate thin shell wormholes was developed. Basically, this methodology consists of taking two copies of a manifold (for example the exterior Schwarzschild solution), and remove from them four dimensional. After that one is left with two geodesically incomplete manifolds with boundaries given by timelike hypersurfaces and, once these hypersurfaces are identified, the resulting space--time is geodesically complete, giving rise to a wormhole throat connecting two asymptotically flat regions. Along this line, dozens of articles have been done \cite{Eiroa:2008ky,Godani2022,Mazharimousavi2022} (and references therein). In a more widely context, beyond Einstein theory, modified gravity theories and higher dimensional ones, wormhole solutions and their properties have been explored. For example, in $f(R)$ gravity theory some interesting results have been obtained in \cite{PhysRevD.80.104012, Harko2013,Capozziello2012,eiroa:2015hrt, PhysRevD.94.044041,Calza:2018ohl,Golchin:2019qch,Lobo:2020vqh} (employing both the metric and Palatini formalism). However, as was stated in \cite{Bronnikov2010} wormhole solutions satisfying energy conditions are forbidden within the framework of $f(R)$ theory, unless states with negative kinetic energy (ghost fields) are propagated. Then, the $f(R)$ gravity scenario has some drawbacks like the GR arena.}

{In order to overcome the above impasse, some authors have developed many interesting wormhole models in another scenarios, such as higher dimensional theories like Einstein--Gauss--Bonnet \cite{Bhawal:1992sz}, warped extra dimensions \cite{Kar2022}, GR extensions such as $f(R,T)$ gravity \cite{Zubair:2021clw}, scale--dependent \cite{Contreras2018}, asymptotically safety gravity \cite{Alencar:2021enh}, conformal Weyl gravity \cite{Lobo2008,Varieschi2016}, teleparalell gravity \cite{Ditta2021}, kinetic gravity \cite{Korolev2020}, hybrid Palatini--metric gravity \cite{hibrid}, action dependent Lagrangian theories \cite{Ayuso2021} and by incorporating 3--form fields \cite{Barros2018} and pole dark energy \cite{Zangeneh2021}, to name a few. In all the mentioned works, the non violation of energy conditions and the satisfaction of the flare--out condition are achieved and justified by extra dimensions and new terms modifying the usual Einstein theory of gravity, GR.}

{As it is well--known, solving Einstein field equations is not a trivial task, even though in the case where spherical symmetry is invoked. To overcome this issue, recently a new methodology called gravitational decoupling by means of minimal geometric deformation (MGD from now on) scheme \cite{Ovalle2017,Ovalle2018} has been suggested. Originally, the MGD methodology was employed in the so--called Brane--World scenario \cite{Ovalle2013,Ovalle2013a,Casadio2012,Casadio2014,Casadio2015,Casadio2015a}. After that, it was used as a gravitational source decoupling scheme, to extend isotropic spherical solution of the Einstein field equations to anisotropic domains \cite{Ovalle2018}. As of now, this approach has been used in different branches to spread out or build up new solutions of the Einstein equations and its extensions.} For example, in considering spherically symmetric perfect fluid matter distributions representing neutron stars, this methodology has been employed to translate these solutions to an anisotropic domains \cite{Ovalle2018,Gabbanelli2018,leond,sanchez,abellan,abellana,leona,rocha,rochaa,Zubair2020,rochab,Zubair2021a,Azmat2021,Azmat2021b,Zubair:2021lgt,Azmat:2022zrv}. Also, this approach has been used to obtain new black hole solutions (hairy and non--hairy) \cite{Ovalle2018c,Contreras2021b,Ovalle2021,Ovalle2021b} and, the properties of some of these solutions have been analyzed in \cite{Ramos2021}. Beyond the mentioned scenarios, this scheme was spread out into the context of 2+1 \cite{Panotopoulos2018, Contreras2018a,Contreras2019} and higher dimensions \cite{Estrada2019,Estrada2019b}, cosmology \cite{linares}, black hole thermodynamics \cite{Estrada2020}, black strings \cite{silva}, holography \cite{tomaz,Rocha2020}, axially symmetric geometries \cite{Sharif2020a}, modified gravity theories \cite{Sharif2020b,Muneer2021,Phys2022} and the so--called inverse problem\footnote{The inverse problem consists in finding the isotropic counterpart of a solution, through its anisotropic version.} \cite{Contreras2018b,Contreras2018c,Contreras2019a}.

One of the most interesting features of this methodology, is the richness of the decoupler sector $\{\theta_{\mu\nu}; f(r)\}$. On one hand, the decoupler sector can be closed using different tools, such as the so--called mimic constraint \cite{Ovalle2018}, interpreting the $\theta$--sector as a scalar field \cite{Ovalle2018a,Arias2022}, using the gravitational complexity factor \cite{Casadio2019,Contreras2021c,Andrade2021,hidalgo,Heras2022}, to name a few. Secondly, as only one metric potential is being deformed (the radial one), there is no exchange of energy between the sources (the seed source and the $\theta$--sector), thus the sources interact only gravitationally. This is so because in the MGD case, Bianchi's identities are satisfied. On the other hand, when both metric potentials are deformed, that is, the so--called extended MGD (e--MGD) \cite{Ovalle2019} (see \cite{silvab,leon,Sharif2020c,Sharif2020d,Sharif2020e,Sharif2020f,Sharif2021,Zubair2021b} for some applications of e--MGD), exists an energy exchange between the interacting sources \cite{Ovalle2022,Contreras2022}, where the deformation function of the temporal metric potential, is playing a major role in this energy exchanging.

In this work, we employ gravitational decoupling by means of MGD, to obtain new wormhole solutions in the GR scenario, where the wormhole seed space--time corresponds to the well--known Morris--Thorne solution \cite{Morris1988,Morris1988A}. To do so, we interpreted the temporal component of the $\theta$--sector as a density profile inspired in non--commutative geometry, Gaussian density profile \cite{Smailagic:2004yy,Nicolini:2005vd,Nicolini:2009gw} and Lorentzian density profile \cite{Mehdipour:2011mc}. This leads to the determination of the $f(r)$ decoupler function and, with information the remaining components of the $\theta$--sector are obtained. Interestingly, by constraining the $\alpha$ parameter, it is possible to obtain solutions with positive density and a small quantity of exotic matter, at the wormhole throat and its neighborhood. Moreover, the models are asymptotically flats and are respecting the flare--out condition, a necessary condition to be a traversable wormhole. 

The first principal reason, to consider the temporal component of the $\theta$--sector to be Gaussian and Lorentzian density profiles, proposed in the context of the noncommutative geometry, lies on the fact of the unknown nature of the $\theta$--sector. As it is well--known, noncommutative could cure divergences that appear in GR, as done in noncommutative quantum field theory\footnote{In such a case, by introducing an exponential cut--off in the Green function of Feynman diagrams at large momenta}\cite{Smailagic:2004yy,Nicolini:2005vd,Nicolini:2009gw}, using the coherent state formalism \cite{Smailagic:2003yb,Smailagic:2003rp} approach. Particularly, divergences appearing as point--like structures (Dirac's delta). So to eliminate this, one can define a suitable smeared functions (Gaussian and Lorentzian distributions for example) in term of noncommutative constants \cite{Smailagic:2004yy,Nicolini:2005vd,Nicolini:2009gw}. Furthermore, the inclusion of these smeared functions does not require the modification of Einstein tensor, thus in makes a suitable condition to interpret the $\theta$--sector (its temporal component) as these density profiles. Although, at the end of the process one ends with a \emph{extended} Einstein tensor coming from a non--usual superposition of the original Einstein tensor and a quasi--Einstein tensor \cite{Ovalle2017}.

The second main reason to match the temporal component of the $\theta$--sector with the mentioned smeared density profiles, lies on the fact that these smeared functions have a monotonically decreasing behavior and are positive defined everywhere. So, the decoupler function $f(r)$ shall be decreasing in nature. This fact is quite relevant, since usually the wormhole seed space--time is asymptotically flat, then this feature is preserved and inherited to the minimally deformed solution\footnote{Of course, asymptotically flat behavior at large enough distances, is not a mandatory condition to have a traversable wormhole. Nevertheless, as this feature is determined by the shape function, and this element is quite involved with the behavior of the density. Then, if the space--time is not asymptotically flat, probably the density and mass of the structure shall not be bounded quantities. Although, this problem can be overcome by using the Israel--Darmois junction conditions, by joining the wormhole space--time with the exterior Schwarzschild solution at some point $r^{*}>r_{0}$ (with $r_{0}$ the wormhole throat).}. Besides, this leads to an effective or total density positive defined everywhere. Usually, in the GR realm is non easy to satisfy at the same footing both conditions, so if the solution is asymptotically flat, then its density is negative and vice--versa. In this sense, the usage of these specific density profiles along with MGD, is leading to an asymptotically flat space--time accompanied with a positive defined density. So, as can be seen the conjunction of the mentioned ingredients, constitutes a good scenario to extend Morris--Thorne wormhole solutions and including new properties.

The article is organized as follows. Sec. \ref{sec2} presents the fields equations, MGD methodology and the strategy to solve the $\theta$--sector. In Sec.\ref{sec3}, the minimally deformed Morris--Thorne wormhole solutions are presented, in conjunction with a detailed geometrical characterization and matter distribution analysis. Finally, Sec. \ref{sec4} concludes the article. 

Throughout the manuscript, the mostly positive metric signature $\{-,+,+,+\}$ is employed and relativistic geometrized units are used \i.e., units where the speed of light $c$ and Newton's gravitational constant are equal to  $c=G=1$. Therefore, the overall coupling constant $\kappa^{2}$ is equal to 8$\pi$.

\section{Fields equations and methodology }\label{sec2}

In this section, we shall discuss in brief how gravitational decoupling by MGD works in the context of GR applied to a spherically symmetric and static space--time. After that, the seed space--time and its energy--momentum tensor is presented and, the setup to close the system of equations corresponding to the $\theta$--sector is given. 

\subsection{Gravitational Decoupling by MGD revisited}

Let us consider Einstein field equations in their general form. These are written as
\begin{equation}\label{eq2}
G_{\mu\nu}\equiv R_{\mu\nu}-\frac{R}{2}g_{\mu\nu}=8\pi \tilde{T}_{\mu\nu}.
\end{equation}
{The starting point is to couple via a dimensionless parameter $\alpha$, a new matter sector $\theta_{\mu\nu}$ to the seed energy--momentum tensor $\tilde{T}_{\mu\nu}$. In this case\footnote{It should be noted that the seed energy--momentum tensor, could be in principle the energy--momentum tensor of an isotropic, charged isotropic, charged anisotropic matter field distribution, to name a few. That is, there is not restriction on the form of $\tilde{T}_{\mu\nu}$. In this occasion we are taking a pure anisotropic matter distribution as a seed space--time, since the seed wormhole space--time is driven by an anisotropic matter distribution (for further details see next sections).   }, we shall take $\tilde{T}_{\mu\nu}$ corresponding to an anisotropic matter distribution
\begin{equation}\label{eq7}
\tilde{T}_{\mu\nu}=\left(\tilde{\rho} +\tilde{p}_{\perp}\right)U_{\mu}U_{\nu}+g_{\mu\nu}\tilde{p}_{\perp}+\left(\tilde{p}_{r}-\tilde{p}_{\perp}\right)\chi_{\mu}\chi_{\nu},
\end{equation}
with $\tilde{\rho}$ is the energy--density, $\tilde{p}_{r}$ and $\tilde{p}_{\perp}$ being the pressures in the principal directions \i.e, the radial and tangential ones, respectively. The four--velocity vector of the above fluid distribution is characterized by the time--like vector $U^{\nu}$. Moreover $\chi^{\nu}$ is a unit space--like vector in the radial direction (orthogonal to $U^{\nu}$).
Henceforth, the right hand side of (\ref{eq2}) becomes
\cite{Ovalle2018}
\begin{equation}\label{eq3}
T_{\mu\nu}=\tilde{T}_{\mu\nu}+\alpha\theta_{\mu\nu},
\end{equation}
where in principle, the new source $\theta_{\mu\nu}$ can be a scalar, vector or tensor field. It is clear that the case $\alpha=0$ yields to the genuine GR theory.  
}

{Next, for the most general spherically symmetric and static line element, representing a wormhole space--time \emph{a la} Morris--Thorne \cite{Morris1988,Morris1988A}
\begin{equation}\label{wormhole}
    ds^{2}=-dt^{2}+\frac{dr^{2}}{1-\frac{\tilde{b}(r)}{r}}+r^{2}d\Omega^{2},
\end{equation}
the field equations (\ref{eq2}) acquire the following form 
\begin{eqnarray}\label{eq4}
8\pi\left(\tilde{\rho}+\alpha\theta^{0}_{\ 0}\right)(r)&=&\frac{\tilde{b}^{\prime}(r)}{ r^{2}}, \\ \label{eq5}
8\pi\left(\tilde{p}_{r}-\alpha\theta^{1}_{\ 1}\right)(r)&=& -\frac{\tilde{b}(r)}{ r^{3}}, \\ \label{eq6}
8\pi\left(\tilde{p}_{\perp}-\alpha\theta^{2}_{\ 2}\right)(r)&=&\frac{\tilde{b}(r)-r\tilde{b}^{\prime}(r)}{2 r^{3}}.
\end{eqnarray}}

{It is clear from equations (\ref{eq4})--(\ref{eq6}), that only the matter sector is being affected by the $\theta$--sector. Furthermore, now we have seven unknowns instead of four as before when $\alpha=0$. In fact, three physical variables $\{\rho(r); p_{r}(r); p_{\perp}(r)\}$, one\footnote{In this case only the so--called shape function $b(r)$ is a geometrical unknown variable. This is so because, we have fixed for the sake of simplicity the temporal metric component, the red--shift $\Phi(r)$, to be zero in order to assure a traversable wormhole solution as was done in \cite{Morris1988,Morris1988A}.} function related with space--time geometry $b(r)$ (see below for further details about the role of this function) and three independent components of $\theta_{\mu\nu}$ \cite{Ovalle2017}. Consequently, these equations form an indefinite system.   }

{So, in view of the above points the question is how to solve this intricate set of equations. One form, is to see the left hand side of (\ref{eq4})--(\ref{eq6}) as effective quantities
\begin{eqnarray}\label{eq8}
{\rho}(r)&\equiv&\left(\tilde{\rho}+\alpha \theta^{0}_{\ 0}\right)(r),\\\label{eq9}
{p}_{r}(r)&\equiv&\left(\tilde{p}_{r}-\alpha \theta^{1}_{\ 1}\right)(r),\\ \label{eq10}
{p}_{\perp}(r)&\equiv& \left(\tilde{p}_{\perp}-\alpha \theta^{2}_{\ 2}\right)(r).
\end{eqnarray}
Then, the system of equations (\ref{eq4})--(\ref{eq6}), could indeed be treated as an anisotropic fluid which would require one to consider four unknown functions, namely the function $b(r)$ and the effective or total quantities given in (\ref{eq8})--(\ref{eq10}). Nevertheless, this is the same as the starting point without the extra field $\theta_{\mu\nu}$. So, as we are interested in obtaining new wormhole solutions (the space--time geometry and the corresponding energy--momentum tensor) starting from an already known solution, we shall take a different approach.}

{The main point here, is to separate new contributions from the known ones. In this case, we want to isolate $\theta_{\mu\nu}$ from $\tilde{T}_{\mu\nu}$. To do so, one needs to associate a geometrical sector to the $\theta_{\mu\nu}$ field. To accomplish it, we shall apply gravitational decoupling by MGD. Essentially, the MGD technique consists in deforming one of the metric potentials in a special way. So, let us consider the following maps
\begin{eqnarray}\label{map1}
\xi(r)\mapsto \Phi(r) &=& \xi(r)+\alpha g(r), \\ \label{map2}
\zeta(r)\mapsto \tilde{b}(r) &=& \zeta(r) +\alpha f(r).
\end{eqnarray}
For the line element (\ref{wormhole}), the red--shift function $\Phi(r)$ is equal to zero, therefore when turning off $\alpha$, one gets $\xi(r)=\Phi(r)=0$. On the other hand, when $\alpha$ is zero the original function $\tilde{b}(r)$ is equal to $\zeta(r)$. With this information at hand and turning on the coupling $\alpha$, the maps (\ref{map1})--(\ref{map2}) become
\begin{eqnarray}\label{map11}
\xi(r)\mapsto \Phi(r) &=&\alpha g(r), \\ \label{map22}
\zeta(r)\mapsto \tilde{b}(r) &=& \zeta(r) +\alpha f(r).
\end{eqnarray}
Therefore, the resulting shape function $b(r)$ is given by
\begin{equation}
    b(r)=\zeta(r)+
    \alpha f(r).
\end{equation}
It is worth mentioning that in the above expressions (\ref{map11})--(\ref{map22}), both decoupler functions $g(r)$ and $f(r)$, only depend on the radial coordinate $r$. This is so because, in this way the spherical symmetry of the seed space--time is preserved.}

{Now, the MGD step means to turn off, that is $g(r)=0$, in this way the effects introduced by $\theta_{\mu\nu}$ lie only on the radial metric function only, encoded in the deformation or decoupler function $f(r)$. So, introducing equation (\ref{map22}) into the set (\ref{eq4})--(\ref{eq6}), one gets
\begin{eqnarray}\label{eq12}
8\pi\tilde{\rho}(r)&=&\frac{\zeta^{\prime}(r)}{ r^{2}}, \\ \label{eq13}
8\pi\tilde{p}_{r}(r)&=& -\frac{\zeta(r)}{ r^{3}}, \\ \label{eq14}
8\pi\tilde{p}_{\perp}(r)&=&\frac{\zeta(r)-r\zeta^{\prime}(r)}{2 r^{3}},
\end{eqnarray}
subject to the following conservation equation
\begin{equation}\label{eq15}
\nabla_{\mu}\tilde{T}^{\mu}_{\ \nu}=0 \rightarrow -\tilde{p}^{\prime}_{r}(r)+\frac{2}{r}\left(\tilde{p}_{\perp}-\tilde{p}_{r}\right)(r)=0.   
\end{equation}
While for the $\theta$--sector one obtains
\begin{eqnarray}\label{eq16}
8\pi\theta^{t}_{\ t}(r)&=&\frac{f^{\prime}(r)}{ r^{2}}, \\ \label{eq17}
8\pi\theta^{r}_{\ r}(r)&=& \frac{f(r)}{ r^{3}}, \\ \label{eq18}
8\pi\theta^{\varphi}_{\ \varphi}(r)&=&\frac{rf^{\prime}(r)-f(r)}{2 r^{3}}.
\end{eqnarray}
Along with the following conservation law
\begin{equation}\label{eq19}
\nabla_{\mu}\theta^{\mu}_{\ \nu}=0 \rightarrow\theta^{\prime r}_{\ r}(r)-\frac{2}{r}\left(\theta^{\varphi}_{\ \varphi}-\theta^{r}_{\ r}\right)(r)=0.  
\end{equation}}

{At this point some comments are pertinent. First, the fact of both parts, namely $\tilde{T}_{\mu\nu}$ and $\theta_{\mu\nu}$ are independently conserved, see equations (\ref{eq16}) and (\ref{eq19}), means that they are related gravitationally. Moreover, it should be noted that the conservation of the $\theta$--sector is a consequence of the conservation of $G_{\mu\nu}$ and $\tilde{T}_{\mu\nu}$. To further understand this point, first  of all consider the situation given by equations (\ref{eq2}) and (\ref{eq3}). In this case Bianchi's identities invoke  
\begin{equation}\label{eq20}
\nabla_{\mu}G^{\mu}_{\ \nu}=0 \Rightarrow \nabla_{\mu}T^{\mu}_{\ \nu}=0,    
\end{equation}
then one has
\begin{equation}\label{eq21}
\nabla_{\mu}\tilde{T}^{\mu}_{\ \nu}+\alpha\nabla_{\mu}\theta^{\mu}_{\ \nu}=0.    
\end{equation}
The zero in the right hand side of the above expression (\ref{eq21}), has two different meanings: (i) $\nabla_{\mu}\tilde{T}^{\mu}_{\ \nu}=\nabla_{\mu}\theta^{\mu}_{\ \nu}=0$ or (ii) $\nabla_{\mu}\tilde{T}^{\mu}_{\ \nu}=-\nabla_{\mu}\theta^{\mu}_{\ \nu}$. The former indicates that each source is covariantly conserved,  and  therefore  the  interaction  between  them  is purely gravitational. The second option, indicates an exchange of energy between these sources (for further details about this point see \cite{Ovalle2022}). In the present situation, we are placed on case (i). What is more, it is not hard to check that equation (\ref{eq19}) is a linear combination of the field equations (\ref{eq16})--(\ref{eq18}).}

{It is worth mentioning that equations (\ref{eq15}) and (\ref{eq19}) are the hydrostatic equation describing the balance of the configuration. As can be seen there is a missing term in both expressions. Specifically, the term proportional to $\Phi^{\prime}$ is absent. Indeed as we have considered $\Phi=0$, this term is not present. However, when the red--shift function is not taken to be zero or constant, then the missing term is restored introducing into the system a gravitational gradient (related with tidal forces).}

{Finally, to close this section it is remarkable to note that the MGD approach turns the  indefinite system (\ref{eq4})--(\ref{eq6}) into the  set of equations for an imperfect fluid $\{\tilde{\rho}(r);\tilde{p}_{r}(r); \tilde{p}_{\perp}; \mu(r)\}$ plus a simpler set of four unknown functions $\{\theta^{t}_{\ t}(r);\theta^{r}_{\ r}(r);\theta^{\varphi}_{\ \varphi}(r);f(r)\}$, where the first set is already solved by the so--called seed space--time, whilst to solve the second system, one can implement several ways. Regarding this last point, the next section is devoted to explain the methodology or strategy to close the $\theta$--sector and thus find new wormhole solutions.   }

\subsection{The strategy}

As we are interested in investigating how gravitational decoupling by means of MGD affects a wormhole space--time, in this case the so--called seed solution is taken to be the well--known Morris--Thorne (MT) solution \cite{Morris1988,Morris1988A}
\begin{equation}\label{morristhorne}
ds^{2}=-dt^{2}+\frac{dr^{2}}{1-\frac{r^{2}_{0}}{r^{2}}}+r^{2}d\Omega^{2},   
\end{equation}
with energy--momentum tensor $\tilde{T}_{\mu\nu}$ given by
\begin{equation}\label{tmunuMT}
\tilde{T}_{\mu\nu}=\frac{r^{2}_{0}}{8\pi r^{4}}\text{diag}\left(-1, -1, 1, 1\right).   
\end{equation}
From the above line element it is clear that the shape $\zeta(r)$ and the red--shift $\Phi(r)$ functions are given by
\begin{equation}
    \zeta(r)=\frac{r^{2}_{0}}{r} \quad \mbox{and} \quad \Phi(r)=0,
\end{equation}
where $r_{0}$ is a positive constant parameter defining the wormhole throat size. So, taking into account the general minimal deformation given by the Eq. (\ref{map11}), the MT space--time (\ref{morristhorne}) acquires the following form
\begin{equation}\label{morristhorne1}
ds^{2}=-dt^{2}+\frac{dr^{2}}{1-\frac{r^{2}_{0}}{r^{2}}-\frac{\alpha f(r)}{r}}+r^{2}d\Omega^{2}.    
\end{equation}
Of course, when $\alpha=0$ in (\ref{morristhorne1}) the solution (\ref{morristhorne}) is recovered. The main aim now, is to determine the decoupler function $f(r)$ associated to the $\theta$--sector. To do this one has (in principle) two ways, the first one is to impose a suitable deformation function $f(r)$ and the second way is by solving the $\theta$--sector, considering some relation among the $\theta_{\mu\nu}$ components or by fixing one of them. In this case we shall employ the second option.

As it is well--known, several proposals have been considered in order to properly close or solve the so--called $\theta$--sector \cite{Ovalle2018,abellan,Ovalle2022}. In the present case we choose a specific form for the temporal component of the $\theta$--sector. Concretely, we are going to assume that $\theta^{t}_{\ t}(r)$ is representing the so--called noncommutative Gaussian energy density $\rho_{G}$ of a static and spherically symmetric smeared and particle--like gravitational source \cite{Smailagic:2004yy,Nicolini:2005vd,Nicolini:2009gw}
\begin{equation}\label{Gauss}
    \rho_{G}=\frac{\mu_{G}}{\left(4\pi \mathbb{M}_{G}\right)^{3/2}}\text{Exp}\left(-\frac{r^{2}}{4\mathbb{M}_{G}}\right),
\end{equation}
and the noncommutative Lorentzian energy density $\rho_{L}$ \cite{Mehdipour:2011mc}
\begin{equation}\label{Lorentz}
    \rho_{L}=\frac{\sqrt{\mathbb{M}_{L}}\mu_{L}}{\pi^{2}\left(\mathbb{M}_{L}+r^{2}\right)^{2}},
\end{equation}
in order to generate two minimally deformed MT wormhole space--times. As mentioned before, these matter profiles have been widely used in the construction of wormhole structures in different gravitational frameworks such as GR \cite{Rahaman2013,Rahaman2015,Kuhfittig2016,Kuhfittig2017}, Gauss--Bonnet \cite{Rani2016}, $f(R)$ \cite{Kuhfittig2018,Kuhfittig2020,Kuhfittig2021} and Rastall \cite{Mustafa:2020kng} theories, to name a few. As matter of fact, profiles (\ref{Gauss}) and (\ref{Lorentz}) are positive defined everywhere and monotonically decreasing functions with increasing $r$. Of course, the positivity upon depends on the signature of the parameters $\mu_{G}$ and $\mu_{L}$. 

{It is worth mentioning that, the noncommutativity of the space--time is an intrinsic feature of the manifold itself, being a consequence of string theory where the coordinates of space--time become non--commutative operators on D--brane \cite{Seiberg:1999vs}. This characteristic, can be expressed as a commutator $[x^{\alpha},x^{\beta}]=i\mathbb{M}^{\alpha\beta}$, being $\mathbb{M}^{\alpha\beta}=\mathbb{M}\,\text{diag}\left(\epsilon_{ij}, \epsilon_{ij}\ldots\right)$ with $\mathbb{M}$ constants and $i,j=1,2,3$ spatial indexes, a skew--symmetric matrix describing the discretization of the
space--time, having dimensions of $\text{length}^{2}$ \cite{Smailagic:2004yy}. Thus, the parameters $\{\mu_{G};\mathbb{M}_{G}\}$ and $\{\mu_{L};\mathbb{M}_{L}\}$ are representing the mass particle--like static
spherically symmetric gravitational source and the noncommutative parameter of the Gaussian and Lorentzian distributions, respectively.}

So, putting together Eqs. (\ref{eq16}) and (\ref{Gauss}) one gets
\begin{equation}\label{f1}
    \theta^{t}_{\ t}(r)=\rho_{G}(r)\Rightarrow f_{G}(r)= 2\mu_{G}\left[\text{Erf}\left(\frac{r}{2\sqrt{\mathbb{M}_{G}}}\right)-\left(\pi\mathbb{M}_{G}\right)^{-1/2}\text{Exp}\left(-\frac{r^{2}}{4\mathbb{M}_{G}}\right)r\right]+C_{G},
\end{equation}
where $C_{G}$ is an integration constant with units of length. Therefore, the MT--Gaussian wormhole space--time is given by
\begin{equation}\label{morristhornegaussian}
ds^{2}=-dt^{2}+\left[1-\frac{r^{2}_{0}}{r^{2}}-\frac{2\alpha\mu_{G}}{r}\left[\text{Erf}\left(\frac{r}{2\sqrt{\mathbb{M}_{G}}}\right)-\left(\pi\mathbb{M}_{G}\right)^{-1/2}\text{Exp}\left(-\frac{r^{2}}{4\mathbb{M}_{G}}\right)r\right]-\frac{\alpha C_{G}}{r}\right]^{-1}dr^{2}+r^{2}d\Omega^{2}.    
\end{equation}
Following the same steps, now equating (\ref{eq16}) with (\ref{Lorentz}) one has the following decoupler function $f(r)$
\begin{equation}\label{f2}
    \theta^{t}_{\ t}=\rho_{L}\Rightarrow f_{L}(r)= \frac{4\mu_{L}}{\pi\left(r^{2}+\mathbb{M}_{L}\right)}\left[\left(r^{2}+\mathbb{M}_{L}\right)\text{arctan}\left(\frac{r}{\sqrt{\mathbb{M}_{L}}}\right)-r\sqrt{\mathbb{M}_{L}}\right]+C_{L},
\end{equation}
being $C_{L}$ an integration constant with units of length. Then, the MT--Lorentzian wormhole is being described by the following line element
\begin{equation}\label{morristhornelorentzian}
ds^{2}=-dt^{2}+\left[1-\frac{r^{2}_{0}}{r^{2}}-\frac{4\alpha\mu_{L}}{\pi \left(r^{2}+\mathbb{M}_{L}\right)r}\left[\left(r^{2}+\mathbb{M}_{L}\right)\text{arctan}\left(\frac{r}{\sqrt{\mathbb{M}_{L}}}\right)-r\sqrt{\mathbb{M}_{L}}\right]-\frac{\alpha C_{G}}{r}\right]^{-1}dr^{2}+r^{2}d\Omega^{2}.    
\end{equation}
Once the decoupler function is obtained, the corresponding $\theta$--sector for each solution, is determined by the field equations (\ref{eq17})--(\ref{eq18}) together with the density 
profile (\ref{Gauss}) for the MT--Gaussian wormhole and (\ref{Lorentz}) for the MT--Lorentzian wormhole model. 

The above results in conjunction with fact that the origin of the $\theta$--sector is unknown \cite{Ovalle2019}, motivated the choice on the temporal component of the $\theta_{\mu\nu}$ source, as the static and spherically symmetric smeared gravitational source (\ref{Gauss}) and (\ref{Lorentz}). As it is well--known, it is not necessary to modify Einstein tensor to include the $\theta$--sector in the GR framework. Of course, via minimal coupling principle between matter and gravity, it can be coupled by introducing a general Lagrangian density and, after taking variations with respect to the metric tensor, terms proportional to $\alpha$ represent the $\theta_{\mu\nu}$ energy--momentum tensor (for further details see \cite{Ovalle2019}). This is in complete agreement with the inclusion of Gaussian and Lorentzian profiles in the GR scenario. Then, in this way the good properties of smeared functions (\ref{Gauss}) and (\ref{Lorentz}), enter in a smooth way enhancing some undesirable issues. Particularly, in the original case the energy density is everywhere negative (\ref{tmunuMT}), then by including and interpreting the component $\theta^{t}_{t}(r)$ as (\ref{Gauss}) and (\ref{Lorentz}), the effective density (\ref{eq8}) shall be positive (at least at the throat and its neighborhood). This is so because, both (\ref{Gauss}) and (\ref{Lorentz}) dominate the behavior of the seed energy density\footnote{Of course, it depends on the choice of the constant parameters involved in expressions (\ref{Gauss}) and (\ref{Lorentz}).}.
On the other hand, profiles (\ref{Gauss}) and (\ref{Lorentz}) are decreasing with increasing radial coordinate. Therefore, the resulting decoupler function $f(r)$ will be too. So, the asymptotically flat behavior at large enough distances of the seminal solution is preserved.

To further support the above results, now we shall explore in general the  asymptotic behavior at $r\rightarrow + \infty$ for the decoupler function $f(r)$ from the general expression (\ref{eq12}) and, the remaining properties will be explored in detail in the next section. So, for cases  (\ref{Gauss}) and (\ref{Lorentz}) the $\theta^{t}_{t}(r)$ component behaves like $\mathcal{O}\left(1/r^{n}\right)$ with $n\in \mathbb{N}$ a large enough number and $\mathcal{O}\left(1/r^{4}\right)$, respectively. Therefore, from equation (\ref{eq12}), one  schematically gets the following behavior for the decopler function $f(r)$ in both cases 
\begin{equation}
    f_{G}(r)\approx D_{G}+\mathcal{O}\left(1/r^{n-3}\right), \quad  f_{L}(r)\approx D_{L}+\mathcal{O}\left(1/r^{3}\right),
\end{equation}
where $D_{G}$ and $D_{L}$ are arbitrary integration constants. So, the asymptotic behavior when $r\rightarrow +\infty$ for the decoupler function in both cases is constant. Then, the asymptotically flat behavior of the seed space--time is preserved. As can be seen, this is very good feature of profiles (\ref{Gauss}) and (\ref{Lorentz}), what is more as we shall see in next section, these properties percolate not only at the level of geometric structure, but also at the level of thermodynamic properties.

\section{The models}\label{sec3}

\subsection{Geometrical study}

As it is well--known, any space--time structure must satisfy certain geometrical requirements to describe a traversable wormhole solution. These conditions contemplate \cite{Morris1988,Morris1988A,Visser:1995cc}
\begin{enumerate}
    \item To avoid horizons, $g_{tt}(r)=e^{\Phi(r)}\neq0$ at $r=r_{0}$. This means that $\Phi(r)$ is everywhere finite.
    \item The throat of the wormhole is located and defined by the condition $b(r_{0})=r_{0}$. This condition makes the metric singular, however, it is just an apparent singularity not a real one.
    \item Both, the red--shift $\Phi(r)$ and shape $b(r)$ functions should be finite in the limit $r\rightarrow +\infty$.
     \item The metric component $g^{-1}_{rr}(r)=1-\frac{b(r)}{r}$ must be always positive  $\forall r \in [r_{*},+\infty)$, where $r_{*}=r_{0}+\epsilon$ with $\epsilon$ a small positive quantity. This is so because at the wormhole throat position, $g^{-1}_{rr}(r_{0})=0$ (the apparent singularity). This is necessary in order to prevent event horizons, where the metric component $g_{rr}(r)$ changes sign.
    \item The so--called flare--out condition 
    \begin{equation}
        \frac{b(r)-rb'(r)}{2b^{2}(r)}>0
    \end{equation}
    and the same one evaluated at the wormhole throat 
    \begin{equation}
        b'(r_{0})<1,
    \end{equation}
    should be respected.
\end{enumerate}

The above list provides the minimum criteria in getting traversable wormhole structures. In this concern, there are some more requirements constraining the red--shift $\Phi(r)$ at the wormhole position and at the spatial station location. Additionally, these conditions are related with tidal accelerations, along the radial and transverse directions. Nevertheless, as we are considering $\Phi(r)=0$ these conditions are automatically satisfied, except the constraint on the acceleration along the radial direction which restricts the traveler speed (for further details see \cite{Morris1988,Morris1988A,Visser:1995cc}).

Now, we are in position to check step by step every point expressed above. So, as the seminal MT solution is being transformed, it is clear that the seed shape function $\mu(r)$ already satisfy point 2. In fact, $\mu(r_{0})=\frac{r^{2}_{0}}{r_{0}}=r_{0}$. Therefore, in general the decoupler or deformation function $f(r)$ must fulfill $f(r_{0})=0$. Thus, from expressions (\ref{f1}) and (\ref{f2}) one needs to enforce the mentioned condition by restricting the integration constants $C_{G}$ and $C_{L}$ for each model respectively. So, for the Gaussian model one gets
from (\ref{f1})
\begin{equation}\label{CG}
f_{G}(r_{0})=0 \Rightarrow C_{G}=-2\mu_{G}\left[\text{Erf}\left(\frac{r_{0}}{2\sqrt{\mathbb{M}_{G}}}\right)-\left(\pi\mathbb{M}_{G}\right)^{-1/2}\text{Exp}\left(-\frac{r^{2}_{0}}{4\mathbb{M}_{G}}\right)r_{0}\right],
\end{equation}
whilst from (\ref{f2}) one obtains
\begin{equation}\label{CL}
f_{L}(r_{0})=0 \Rightarrow C_{L}=-\frac{4\mu_{L}}{\pi\left(r^{2}_{0}+\mathbb{M}_{L}\right)}\left[\left(r^{2}_{0}+\mathbb{M}_{L}\right)\text{arctan}\left(\frac{r_{0}}{\sqrt{\mathbb{M}_{L}}}\right)-r_{0}\sqrt{\mathbb{M}_{L}}\right].
\end{equation}
Plugging the above results in the corresponding shape function for each model, these acquire the following form:\\

\textbf{Gaussian shape function}

\begin{equation}\label{eq34}
    b_{G}(r)=\frac{r^{2}_{0}}{r}+2\alpha\mu_{G}\left[\text{Erf}\left(\frac{r}{2\sqrt{\mathbb{M}_{G}}}\right)-\left(\pi\mathbb{M}_{G}\right)^{-1/2}\text{Exp}\left(-\frac{r^{2}}{4\mathbb{M}_{G}}\right)r-\text{Erf}\left(\frac{r_{0}}{2\sqrt{\mathbb{M}_{G}}}\right)+\left(\pi\mathbb{M}_{G}\right)^{-1/2}\text{Exp}\left(-\frac{r^{2}_{0}}{4\mathbb{M}_{G}}\right)r_{0}\right],
\end{equation}

\textbf{Lorentzian shape function}

\begin{equation}\label{eq35}
    b_{L}(r)=\frac{r^{2}_{0}}{r}+\frac{4\alpha\mu_{L}}{\pi}\left[\text{arctan}\left(\frac{r}{\sqrt{\mathbb{M}_{L}}}\right)-\frac{r\sqrt{\mathbb{M}_{L}}}{\left(r^{2}+\mathbb{M}_{L}\right)}-\text{arctan}\left(\frac{r_{0}}{\sqrt{\mathbb{M}_{L}}}\right)+\frac{r_{0}\sqrt{\mathbb{M}_{L}}}{\left(r^{2}_{0}+\mathbb{M}_{L}\right)}\right].
\end{equation}

As point 3 suggests, it is desirable to have a finite shape function $b(r)$ when $r\rightarrow+\infty$. This is so because, a finite form function at enough large distances leads to an asymptotically flat space--time since
\begin{equation}
    \lim_{r\rightarrow +\infty} \frac{b(r)}{r}\rightarrow 0.
\end{equation}
Furthermore, the above condition provides the mass $m(r)$ of the wormhole structure is also finite (this point will be clear soon). Besides, as $m(r)$ is a positive quantity, then $b(r)$ should be a positive defined function everywhere. However, if $b(r)$ is not finite when $r\rightarrow+\infty$, consequently the mass $m(r)$ is not finite too. In such a case, we can cut the wormhole space--time at some point $a>r_{0}$ and paste it with the outer Schwarzschild solution \cite{Visser:1995cc,Poisson:1995sv} to have an asymptotically flat space--time with a positive bounded mass $M$ ($M$ is the value of the mass function $m(r)$ when $r\rightarrow +\infty$). In a more widely context, the asymptotically flat behavior of the solution can be relaxed, admitting a different topology of the  space--times connected by the throat. For instance, in \cite{Roman:1992xj} throats connecting asymptotically de Sitter spaces were obtained.

Since the seed space--time is already asymptotically flat with a bounded and positive mass, it cannot be warranted that after applying the MGD method, the resulting space--time preserves such a structure. Evidently, this issue strongly depends on how the $\theta$--sector is closed in order to get $f(r)$, since different strategies yield to different decoupler functions. Therefore, the functional dependency of $f(r)$ with respect to $r$ cannot be predicted, unless the deformation function $f(r)$ is imposed by hand. Then an asymptotically flat behavior depends upon on the form of $f(r)$. On the other hand, the only thing that can be ensured is the spherical symmetry of the final solution, by allowing the deformation function $f(r)$ to depend only on the radial coordinate $r$.

From expressions (\ref{f1}) and (\ref{f2}), it is not hard to check that both $f_{G}(r)$ and $f_{L}(r)$ are finite at enough large distances. In fact

\begin{equation}
    \lim_{r\rightarrow +\infty} f_{G}(r) \rightarrow 2\mu_{G} +C_{G}
\end{equation}
subject to $\mu_{G}\in \mathbb{R}$ and $\sqrt{\mathbb{M}_{G}}>0$ and,
\begin{equation}
    \lim_{r\rightarrow +\infty} f_{L}(r) \rightarrow 2\mu_{L} +C_{L}
\end{equation}
with $\mu_{L}\in \mathbb{R}$. Hence both space--times (\ref{morristhornegaussian}) and (\ref{morristhornelorentzian}) are asymptotically flat in the limit $r\rightarrow +\infty$.

\begin{figure}[H]
\centering
\includegraphics[width=0.33\textwidth]{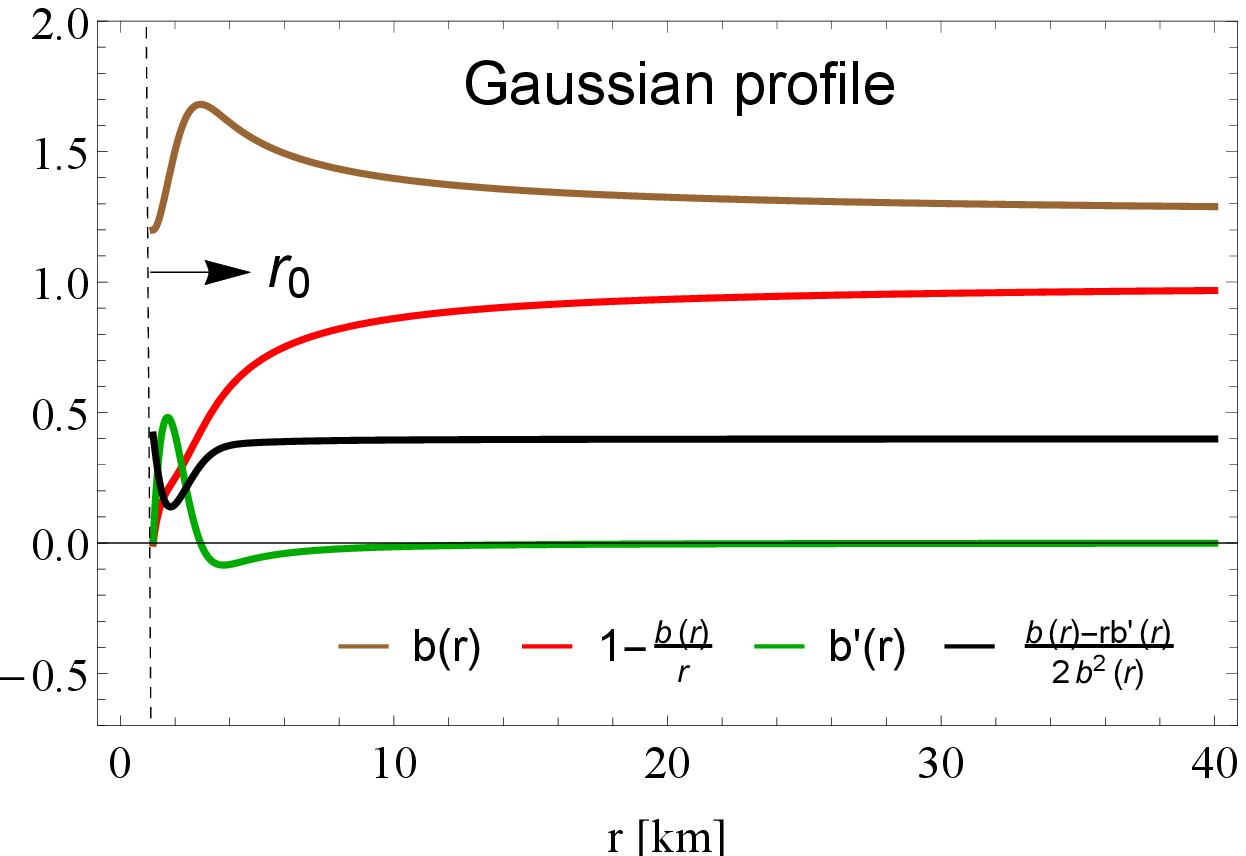}  \ 
\includegraphics[width=0.33\textwidth]{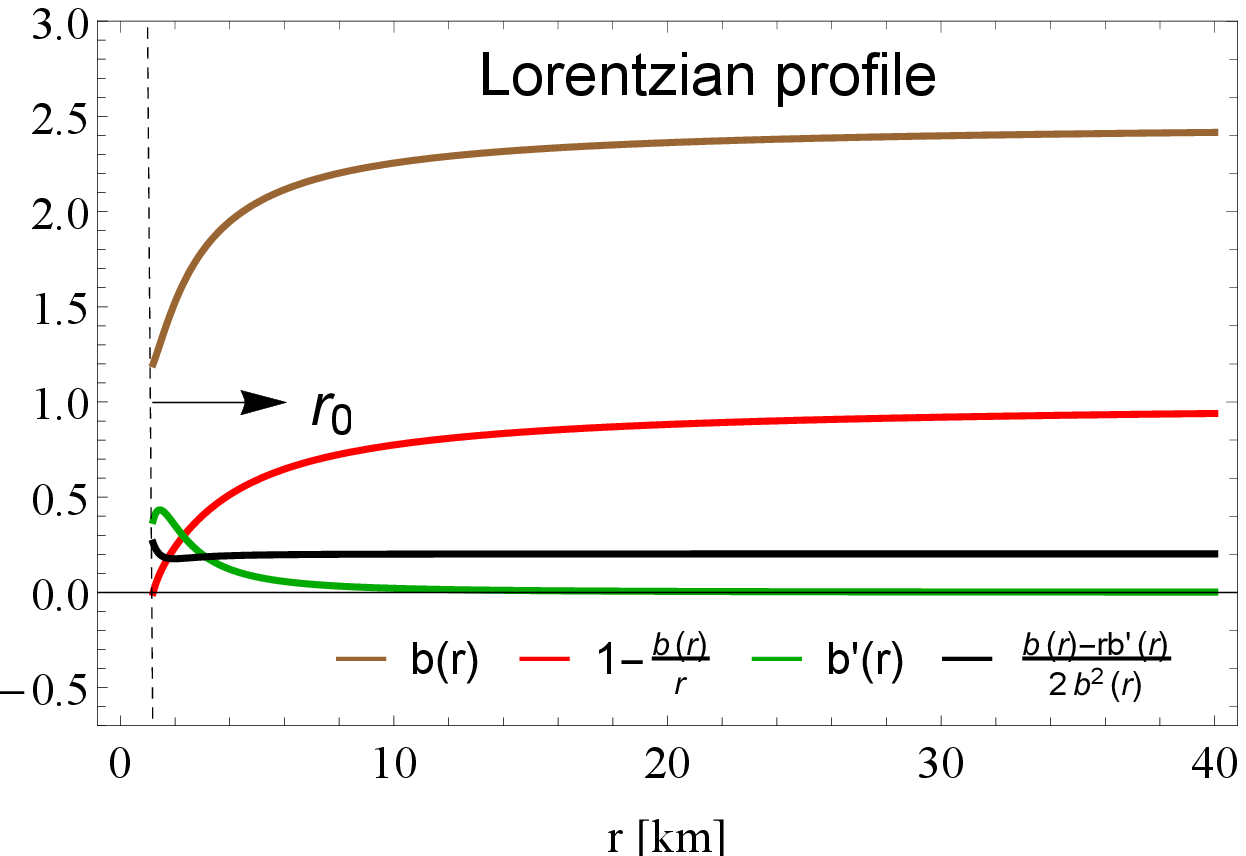}
\caption{Geometrical description for each model, \textbf{Left panel}: Gaussian profile and \textbf{Right panel}: Lorentzian profile, respectively. To build up these plots, numerical values for constant parameters characterizing each model were considered as displayed in table \ref{table2}.
}\label{fig1}
\end{figure}

\begin{figure}[H]
\centering
\includegraphics[width=0.33\textwidth]{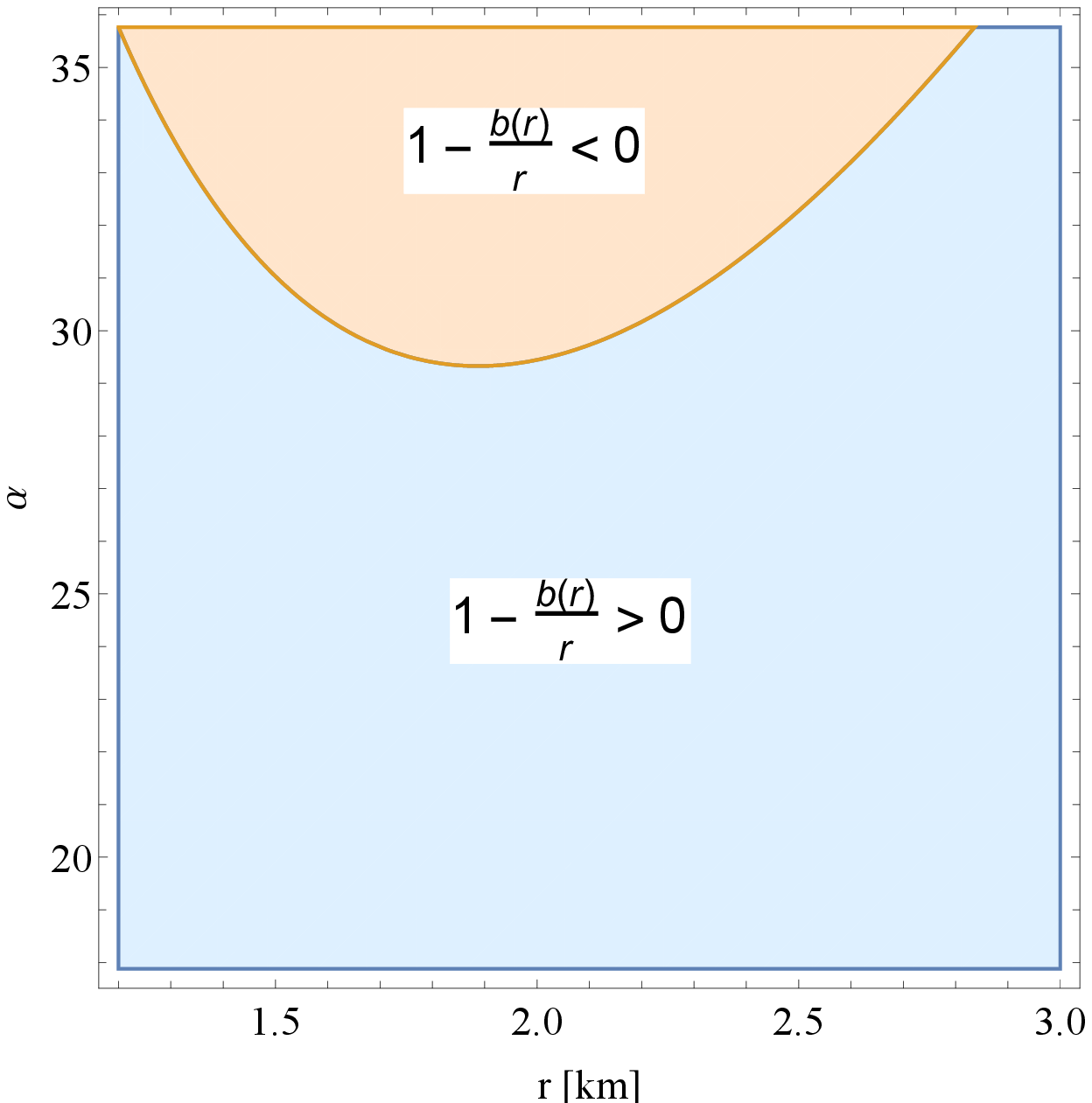}  \ 
\includegraphics[width=0.33\textwidth]{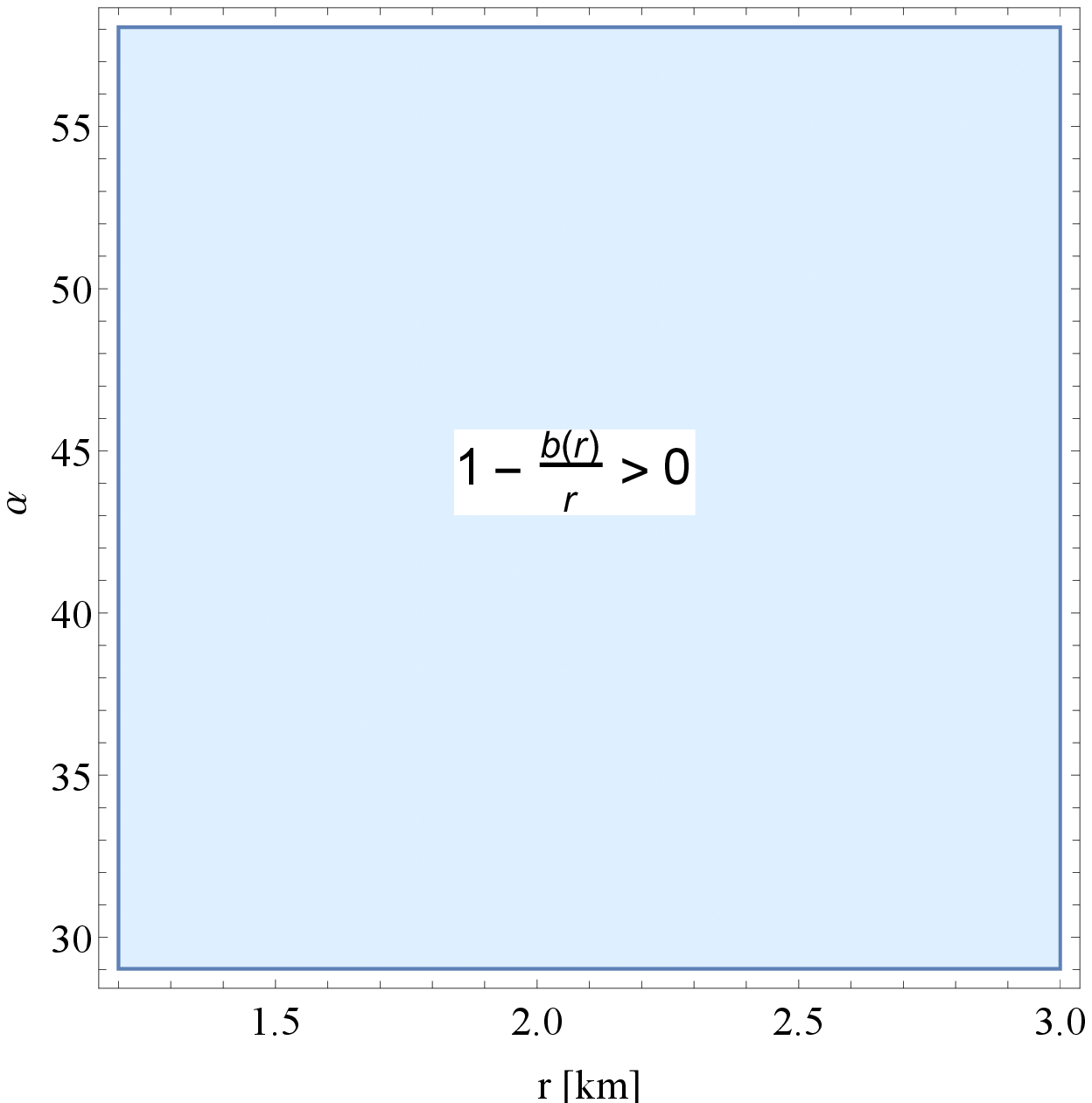}
\caption{Region validity (blue region) for the metric potential $g^{-1}_{rr}$. \textbf{Left panel}: Gaussian model and \textbf{Right panel}: Lorentzian model. Possible values for $\alpha$ in order to avoid signature sign changes, were obtained by using numerical values in table \ref{table2}.
}\label{fig2}
\end{figure}

The mentioned behavior of the solutions can be corroborated in Fig. \ref{fig1}, where in both cases $b_{G}(r)$ and $b_{L}(r)$ are positive 
and finite $\forall r \in [r_{0},+\infty)$. Furthermore, the trend of the $g^{-1}_{rr}(r)$ metric potential is displayed (red line). As can be appreciated, its behavior shows that the radial component of the metric tends to Minkowski space--time coefficient, that is 1, when $r\rightarrow +\infty$. Besides, in both cases $g^{-1}_{rr}(r)$ (consequently $g_{rr}(r)$) is positive everywhere. This is an important point in the wormhole study, since the Lorentzian signature, in this case $(-,+,+,+)$, must be always preserved. Of course, as said before the MGD involves certain changes in the space--time structure, thus the positivity of the radial metric potential of the deformed wormhole depends on some factors such as the values and signature of the constant $\alpha$. In this concern, for chosen values of the parameters $\{\mu_{G};\mathbb{M}_{G}\}$ and $\{\mu_{L};\mathbb{M}_{L}\}$ and the magnitude of the wormhole throat $r_{0}$ (see below and table 1). Fig. \ref{fig2} illustrates the admissible spectrum of $\alpha$, as can be seen for the Gaussian model (left panel) there is a forbidden region where the $g^{-1}_{rr}(r)$ metric potential reverses its signature \i.e., $g^{-1}_{rr}(r)<0$, hence the space--time metric signature becomes $(-,-,+,+)$. Therefore, all these values of $\alpha$ cannot be considered. On the other hand, for the Lorentzian model (right panel) all values for the parameter $\alpha$ are admissible. 

To further corroborate the asymptotically flat behavior of the solutions, we analyze the so--calling embedding diagram. So, from the general line element (\ref{morristhorne}) one can get relevant information about the form of the wormhole structure \cite{Morris1988,Morris1988A,Visser:1995cc}. Concentrating on an equatorial plane, $\theta=\pi/2$, the solid angle element $d\Omega^{2}$ reduces to
\begin{equation}
d\Omega^{2}=d\phi^{2}.
\end{equation}
Moreover, by fixing $t=\text{constant}$, the line element (\ref{morristhorne}) becomes
\begin{equation}\label{redu}
ds^{2}=\frac{dr^{2}}{1-\frac{b(r)}{r}}+r^{2}d\phi^{2}.    
\end{equation}
To visualize the equatorial plane as a 2--dimensional surface encrusted in an 3--dimensional Euclidean space, it is appropriate to consider cylindrical coordinates \begin{equation}
ds^{2}=dz^{2}+dr^{2}+r^{2}d\phi^{2},    
\end{equation}
or equivalently,
\begin{equation}\label{cili}
ds^{2}=\left[1+\left(\frac{dz}{dr}\right)^{2}\right]dr^{2}+r^{2}d\phi^{2}.    
\end{equation}
Contrasting (\ref{redu}) and (\ref{cili}), one gets
\begin{equation}\label{embe}
\frac{dz(r)}{dr}=\pm\left(\frac{r}{b(r)}-1\right)^{-1/2},    
\end{equation}
where the function $z=z(r)$ defines the embedded surface \cite{Morris1988,Morris1988A,Visser:1995cc}. As $b(r)$ is finite in the limit $r\rightarrow +\infty$, 
the expression (\ref{embe}) goes to
\begin{equation}
    \frac{dz(r)}{dr}\bigg|_{r\rightarrow+\infty}=0.
\end{equation}
This tell us that the wormhole space--time is covering two asymptotically flat patches. However, in this case expression (\ref{embe}) might lead to a different scenario. So, taking into account (\ref{map11}) one has
\begin{equation}\label{embe1}
    \frac{dz(r)}{dr}=\pm \left(\frac{r}{\zeta(r)+\alpha f(r)}-1\right)^{-1/2}.
\end{equation}
Then the satisfaction of result (\ref{embe}) upon depends on some conditions. The most general scenarios are:
\begin{itemize}
    \item $f(r)$ is a function that goes faster than $r$ to infinity (polynomial of degree $n>2$ in the variable $r$ or exponential function of $r$, etc.). In such a case (\ref{embe1}) provides $-i$ (imaginary unit), being clear that the space--time topology does not coincide with the Minkowski space--time in the long--distance regime.
    \item $f(r)$ is a polynomial function of degree $n<1$ on the variable $r$, a logarithmic function, for example. In this case equation (\ref{embe1}) goes to $dz/dr\rightarrow0$. Therefore the space--time is asymptotically flat. 
    \item Another interesting case, occurs when the function $f(r)$ is a linear polynomial expression in $r$. In such a case, equation (\ref{embe1}) gives $\alpha$ when $r\rightarrow + \infty$. Therefore the space--time has a topological defect, a solid angle deficit (or excess).  
\end{itemize}
Of course, above we have considered some particular situation about the form of the decoupler function $f(r)$, since more complex or simple expression can be obtained. Nevertheless, with the previous discussion is enough to guess the topological shape of the space--time when the radial coordinate tends to infinite.

\begin{figure}[H]
\centering
\includegraphics[width=0.32\textwidth]{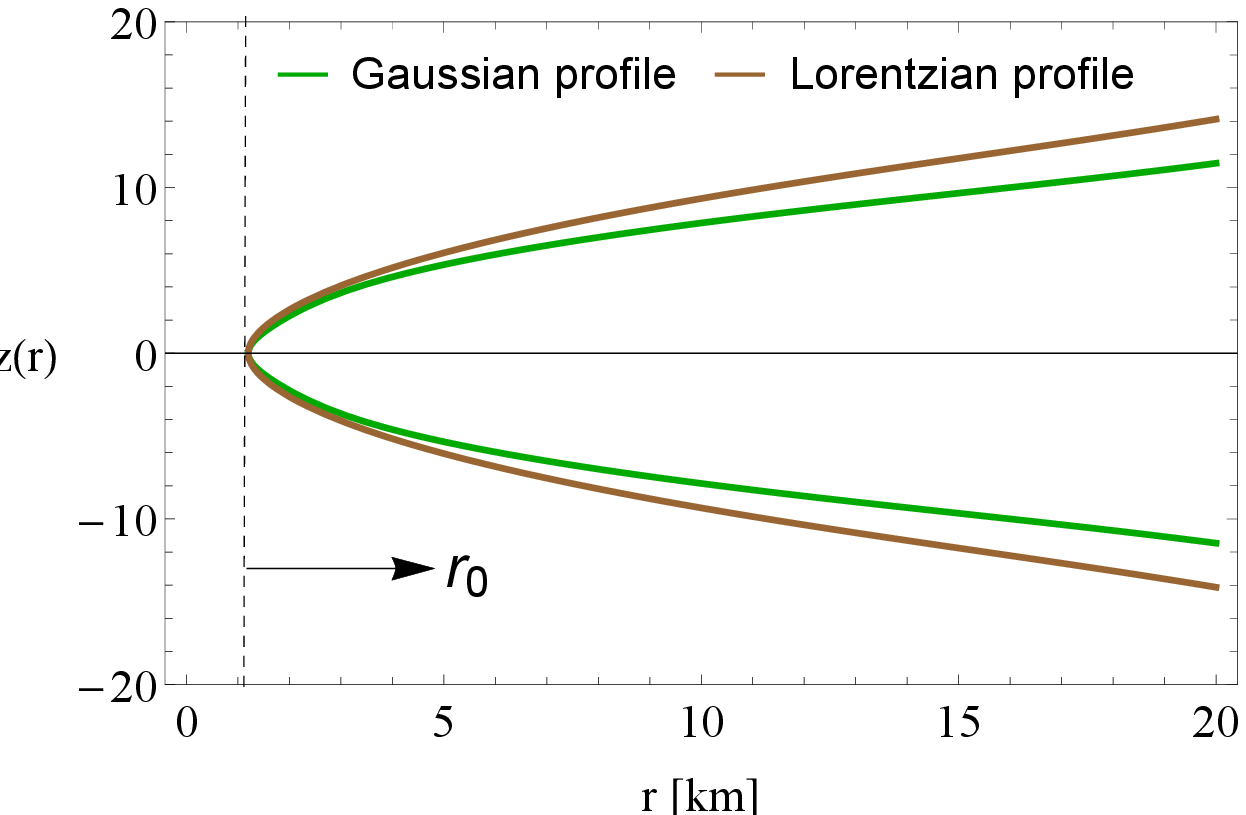}  \ 
\includegraphics[width=0.32\textwidth]{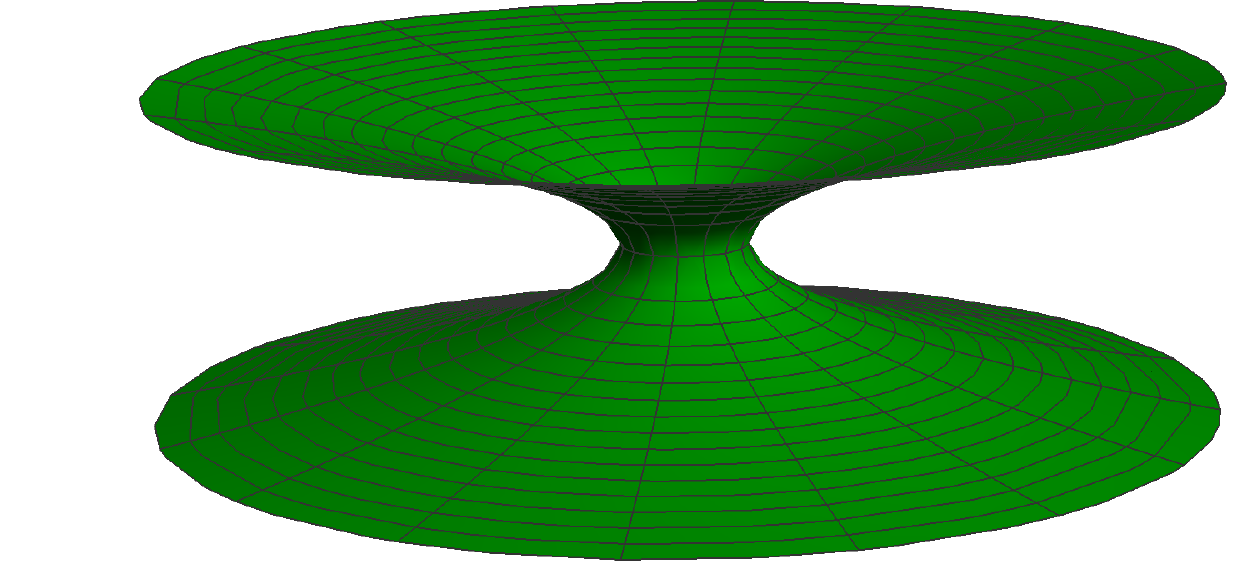} \
\includegraphics[width=0.32\textwidth]{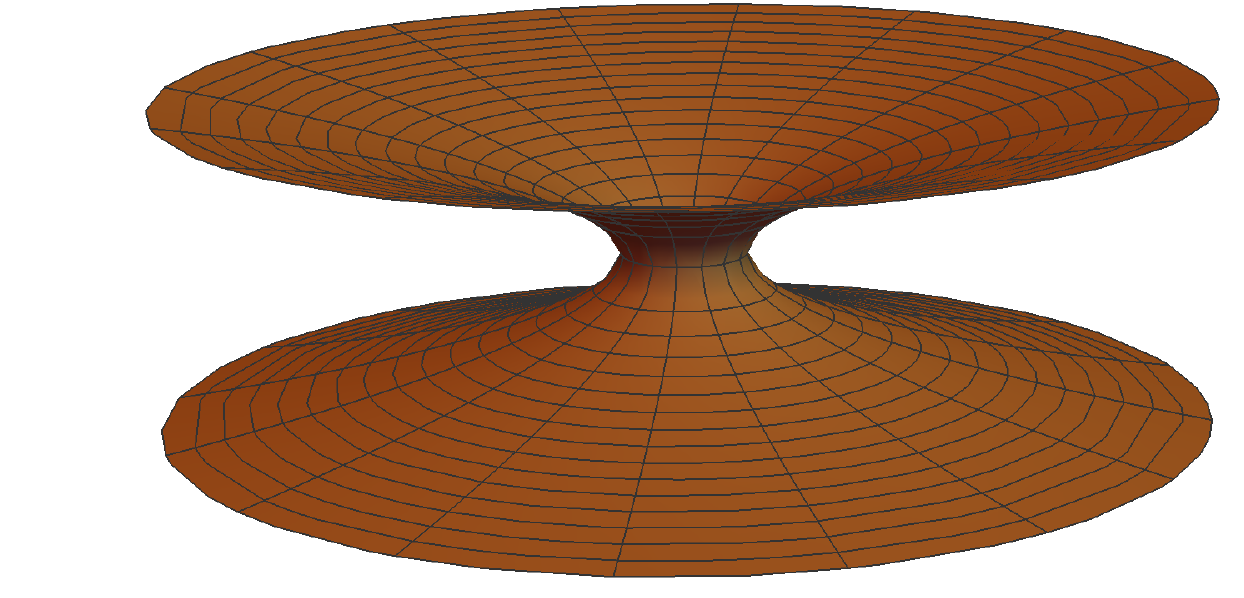}
\caption{ \textbf{Left panel}: 2--dimensional embedding diagram $z(r)$. \textbf{Middle panel}: 3--dimensional hyper--surface for the Gaussian model. \textbf{Right panel}: 3--dimensional hyper--surface for the Lorentzian model. These 2 and 3--dimensional embedding diagrams were constructed by using numerical data showed in table \ref{table2}.
}\label{fig3}
\end{figure}

\begin{figure}[H]
\centering
\includegraphics[width=0.5\textwidth]{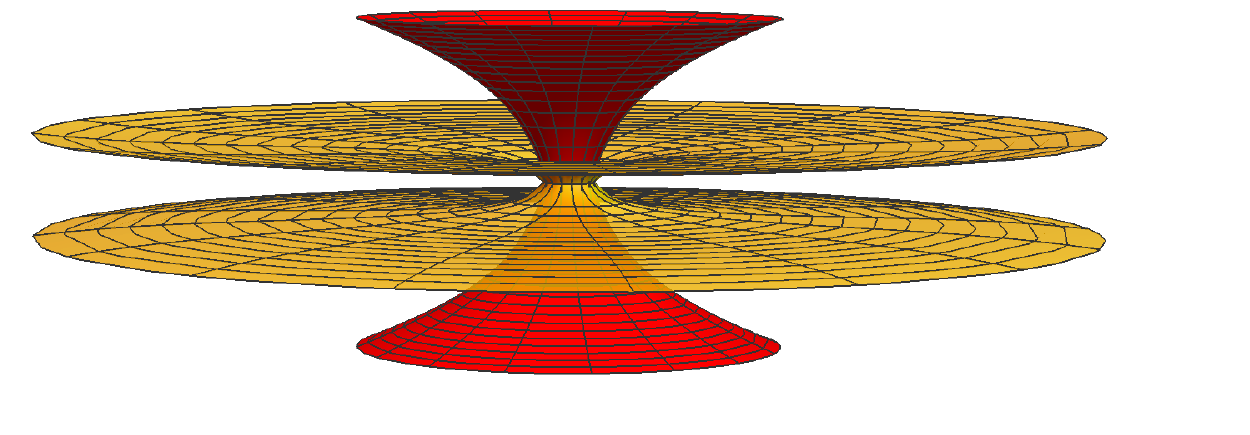}
\caption{ 3--dimensional embedding diagram comparison between original MT space--time and Gaussian profile model. 
}\label{fig6}
\end{figure}

It should be noted that the integration of Eq. (\ref{embe1}) cannot be done analytically. Therefore, a numerical treatment has been done in order to illustrate the wormhole shape. The left panel of Fig. \ref{fig3} is showing 2--dimensional embedding diagram (the $z(r)$ function) versus the radial coordinate $r$ for each model. While middle and right panels are displaying the characteristic form of asymptotically flat wormholes as a 3--dimensional hyper--surface. Furthermore, Fig. \ref{fig6} is depicting a comparison between the 3--dimensional embedding diagram of the MT seed space--time and minimally deformed MT solution for Gaussian density profile (an example). As can be observed, terms proportional to $\alpha$ coupling, are modifying the proper length $l(r)$ of the wormhole throat.

Another relevant aspect from the geometric point of view, is the flare--out condition. When this condition is met, it tells us that the structure does not shrink. Inverting expression (\ref{embe1}) to obtain $dr$ in terms of $dz$ and taking the second derivative one can obtain
\begin{equation}\label{secondderi}
 \frac{d^{2}r(z)}{dz^{2}}=\frac{\zeta(r)+\alpha f(r)-\left[\zeta^{\prime}(r)+\alpha f^{\prime}(r)\right]r}{2\left[\zeta(r)+\alpha f(r)\right]^{2}}>0,  
\end{equation}
This condition should be satisfied at or near the wormhole throat $r=r_{0}$. Specially, at $r_{0}$ expression (\ref{secondderi}) provides
\begin{equation}\label{eq47}
   \frac{d^{2}r(z)}{dz^{2}}\bigg{|}_{r=r_{0}}= b'(r_{0})<1 \Rightarrow \zeta'(r_{0})+\alpha f'(r_{0})<1, 
\end{equation}
where the condition $f(r_{0})=0$ have been used. As can be seen, the flare--out condition evaluated at the wormhole throat is modified by the decoupler function (its first derivative). So, expression (\ref{eq47}) serves to constraint the magnitude and signature of the coupling constant $\alpha$. Indeed, plugging in (\ref{eq47}) expressions (\ref{eq34}) and (\ref{eq35}) for each model respectively, one gets the following bound on the parameter $\alpha$ 
\begin{equation}\label{flare1}
    b'_{G}(r)=-\frac{r^{2}_{0}}{r^{2}}+\alpha\mu_{G}\text{Exp}\left(-\frac{r^{2}}{4\mathbb{M}_{G}}\right)\frac{r^{2}}{\sqrt{\pi}\beta^{3/2}_{G}}<1\Rightarrow \alpha<2\sqrt{\frac{\text{Exp}\left(\frac{r^{2}_{0}}{2\mathbb{M}_{G}}\right)\beta^{3}_{G}\pi}{r^{4}_{0}\mu^{2}_{G}}},
\end{equation}
for the Gaussian case and
\begin{equation}\label{flare2}
    b'_{L}(r)=-\frac{r^{2}_{0}}{r^{2}}+\frac{8\alpha\mu_{L}\sqrt{\mathbb{M}_{L}}\,r^{2}}{\pi\left(r^{2}+\mathbb{M}_{L}\right)^{2}}<1\Rightarrow \alpha<\frac{\pi}{4\mu_{L}}\sqrt{r^{4}_{0}\beta^{-1}_{L}+4r^{2}_{0}+6\mathbb{M}_{L}+4r^{-2}_{L}\beta^{2}_{L}+r^{-4}_{0}\beta^{3}_{L}},
\end{equation}
for the Lorentzian model. As it is observed an upper bound for $\alpha$ has been obtained by evaluating the flare--out condition at the wormhole throat. 

On the other hand, from the field equation (\ref{eq4}) it is clear that the density $\rho(r)$ depends on the first derivative of the shape function. So, if the form function is an increasing function with increasing radial coordinate $r$, its first derivative will be positive, otherwise if the shape function is a decreasing function within the interval $[r_{0},+\infty)$, then its derivative shall be negative. This simple mathematical analysis has a great impact on the behavior of the matter content of the model (see next section), because if $b'(r)<0$ then $\rho(r)<0$. As discussed earlier, in the realm of wormhole solutions, violation of energy conditions mainly occurs due to $p_{r}(r)<0$ and $|p_{r}(r)|>|\rho(r)|$. Hence, if both $\rho(r)$ and $p_{r}(r)$ are negative, the situation is worst. As it is well--known, classical states allowing a negative density are forbidden. Of course, being the wormhole subject a purely hypothetical field, in principle one can allow these kind of matter distributions, although it is desirable to prevent them. In the GR case, sometimes it is not possible to have an asymptotically flat wormhole space--time driven by a positive defined density, neither at the wormhole throat and its vicinity and beyond it. Albeit the non--asymptotically flat behavior can be easily addressed by cutting and pasting the solution at some point $a>r_{0}$ with the outer Schwarzschild solution.

In the present case, the seed shape function, that is, the MT form function $\zeta(r)=\frac{r^{2}_{0}}{r}$ is a decreasing function as $r$ increases, then $\rho(r)<0$ throughout the space--time. However, the introduction of the MGD scheme in principle might help us to get away this issue. In fact, now total or effective density of the system is equal to the combination $\zeta'(r)+\alpha f'(r)$, so if one imposes the constraint $\zeta'(r)<\alpha f'(r)$ subject to $\alpha f'(r)>0$, then $\rho(r)+\alpha \theta^{t}_{\,t}(r)>0$. It is evident that this requirement strongly depends on the behavior (increasing/decreasing) of the decoupler function $f(r)$ and its magnitude with respect to $\zeta'(r)$. Therefore, to guarantee the above statement one needs to take under control the behavior of $\alpha f(r)$, since it is not possible to guess the form of $f(r)$ before solving the $\theta$--sector (independently of the strategy), unless $f(r)$ is being imposed by hand from the very beginning. Accordingly, to overcome this impasse the constant $\alpha$ should be also constrained from below by the flare--out condition evaluated at the throat of the configuration. Therefore, demanding $b'(r_{0})>0$ one gets the following lower bound on $\alpha$ for each model,
\begin{equation}\label{b1}
   \alpha>\sqrt{\frac{\text{Exp}\left(\frac{r^{2}_{0}}{2\mathbb{M}_{G}}\right)\beta^{3}_{G}\pi}{r^{4}_{0}\mu^{2}_{G}}},
\end{equation}
and 
\begin{equation}\label{b2}
    \alpha>\frac{\pi}{8\mu_{L}}\sqrt{r^{4}_{0}\beta^{-1}_{L}+4r^{2}_{0}+6\mathbb{M}_{L}+4r^{-2}_{L}\beta^{2}_{L}+r^{-4}_{0}\beta^{3}_{L}},
\end{equation}
for the Gaussian and Lorentzian model, respectively. The flare--out trend for each model taking into account values displayed in table \ref{table2}, is depicted in Fig. \ref{fig1}. In both cases this condition is satisfied at the throat and beyond it (see green and black lines on the panels).

\begin{table}[H]
\caption{Numerical values of constant parameters for each model. }
\label{table2}
\centering
\begin{tabular}{|c|c|c|c|c|}
\hline
Density Profile & $\mu_{i}$ [km], $(i=G, L)$ & $\mathbb{M}_{i}$ [$\text{km}^{2}$], (i= G, L) & $r_{0}$ [km] & $\alpha$ \\ \hline
Gaussian        &     0.05                     &    0.5                                      & 1.2          & 18         \\ \hline
Lorentzian      &   0.05                        & 0.5                                         & 1.2          & 40          \\ \hline
\end{tabular}
\end{table}

\subsection{Matter distribution analysis}

In this section, we are going to discuss the impact of gravitational decoupling by MGD on energy conditions. Before accounting for the effect on each model, it is instructive to check how the null energy condition (NEC) is in general modified and whether it brings out new insights with respect to the pure GR case. So, from the field equations (\ref{eq4}) and (\ref{eq5}) one has
\begin{equation}
    8\pi\left[\tilde{\rho}(r)+\alpha\theta^{t}_{\,t}(r)+\tilde{p}_{r}(r)-\alpha\theta^{r}_{\, r}(r)\right]=\frac{1}{r^{3}}\left[r\zeta'(r)+\alpha rf'(r)-\zeta(r)-\alpha f(r)\right],
\end{equation}
where evaluating at $r=r_{0}$ the above expression provides
\begin{equation}\label{nec}
 8\pi\left[\tilde{\rho}(r)+\alpha\theta^{t}_{\,t}(r)+\tilde{p}_{r}(r)-\alpha\theta^{r}_{\, r}(r)\right]\bigg{|}_{r=r_{0}}=-\frac{1}{r^{2}_{0}}\left[1-\zeta'(r_{0})-\alpha f'(r_{0})\right],
\end{equation}
where terms on the right member inside the bracket, are just the same ones appearing at the right hand side of expression (\ref{secondderi}). Therefore, the full bracket is positive. Consequently the full right member of (\ref{nec}) is negative. Then, the main point here is that the violation of NEC is not affected by the introduction of the MGD scheme. This is so because $f(r_{0})=0$, thus the full radial pressure at the wormhole throat is being dominated only by the seed shape function $\zeta(r)$ and, despite the deformation function is quite involved in the flare--out condition, in order to respect it $\alpha$ must be constrained, otherwise it is not possible to maintain the wormhole throat open. Naturally, one wonders what has been gained by minimally deforming a solution that by construction violates NEC, if the deformed solution also does. This question shall be answered in short.

\begin{figure}[H]
\centering
\includegraphics[width=0.32\textwidth]{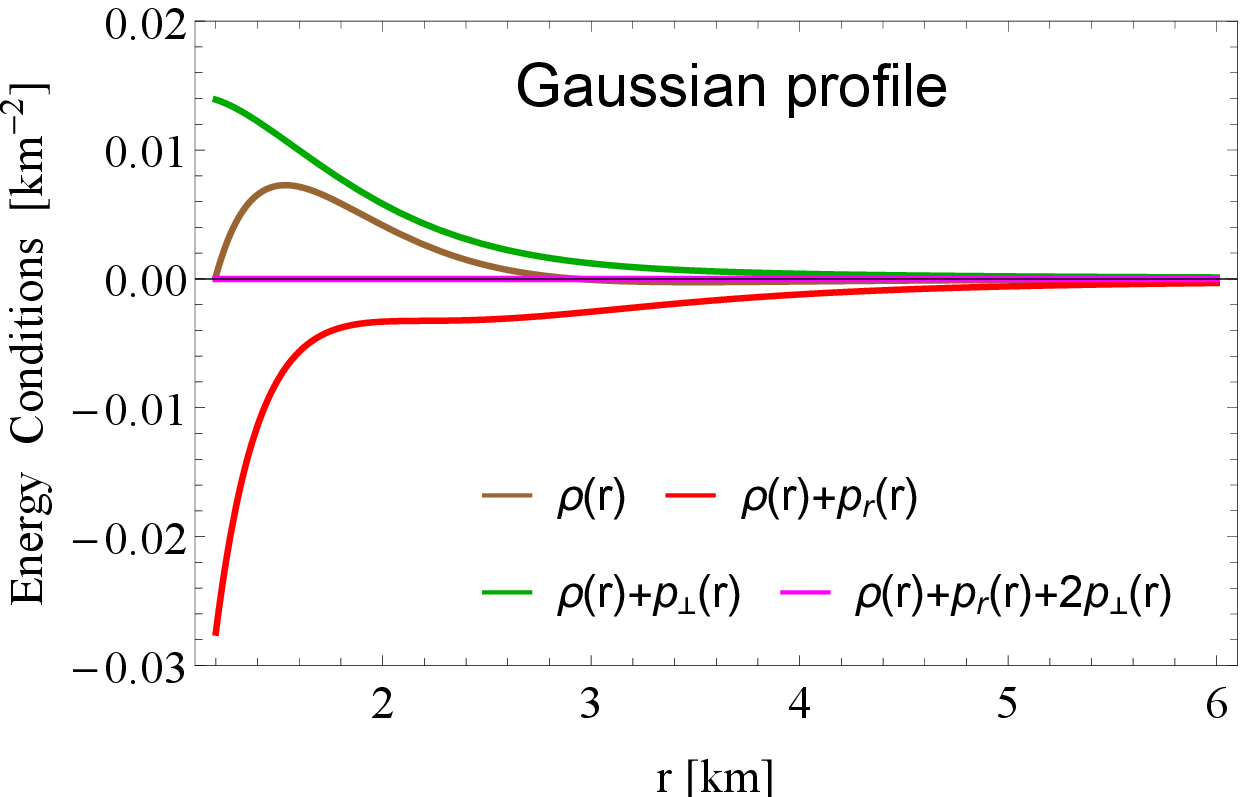}  \ 
\includegraphics[width=0.32\textwidth]{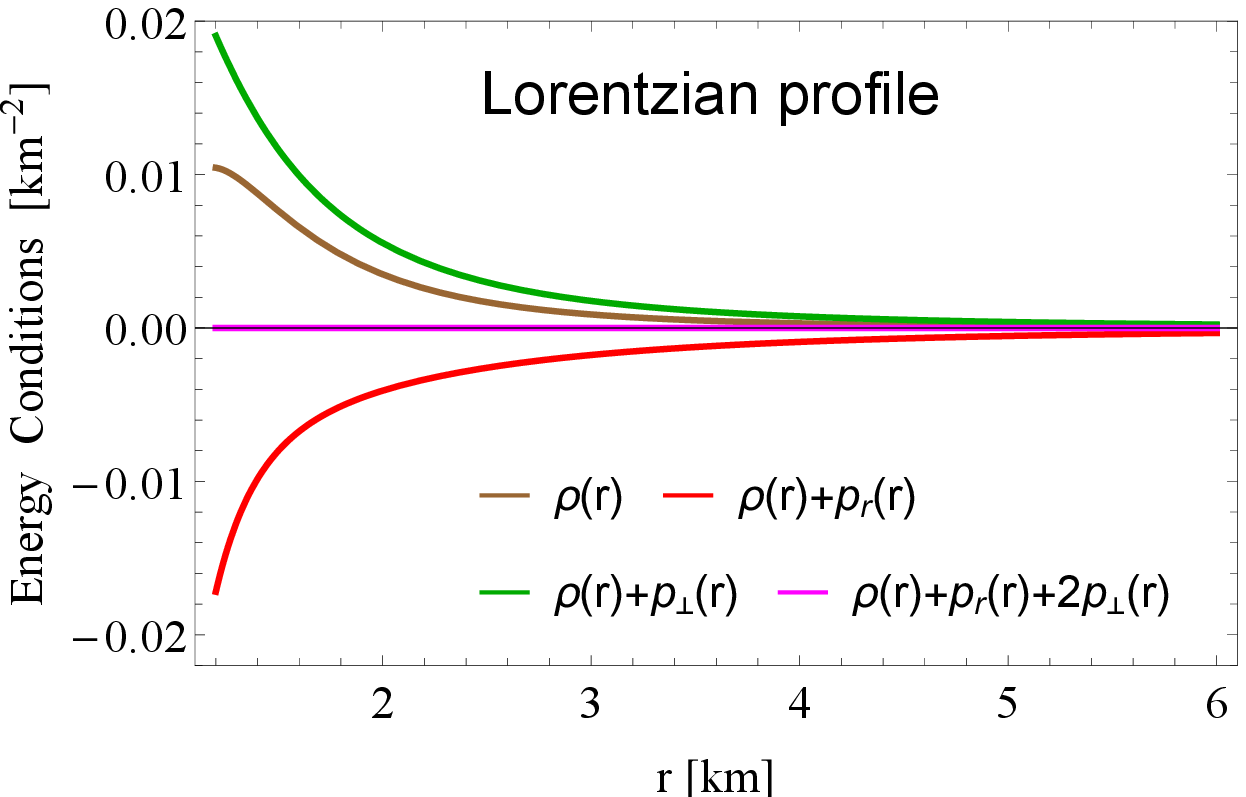} \\
\includegraphics[width=0.32\textwidth]{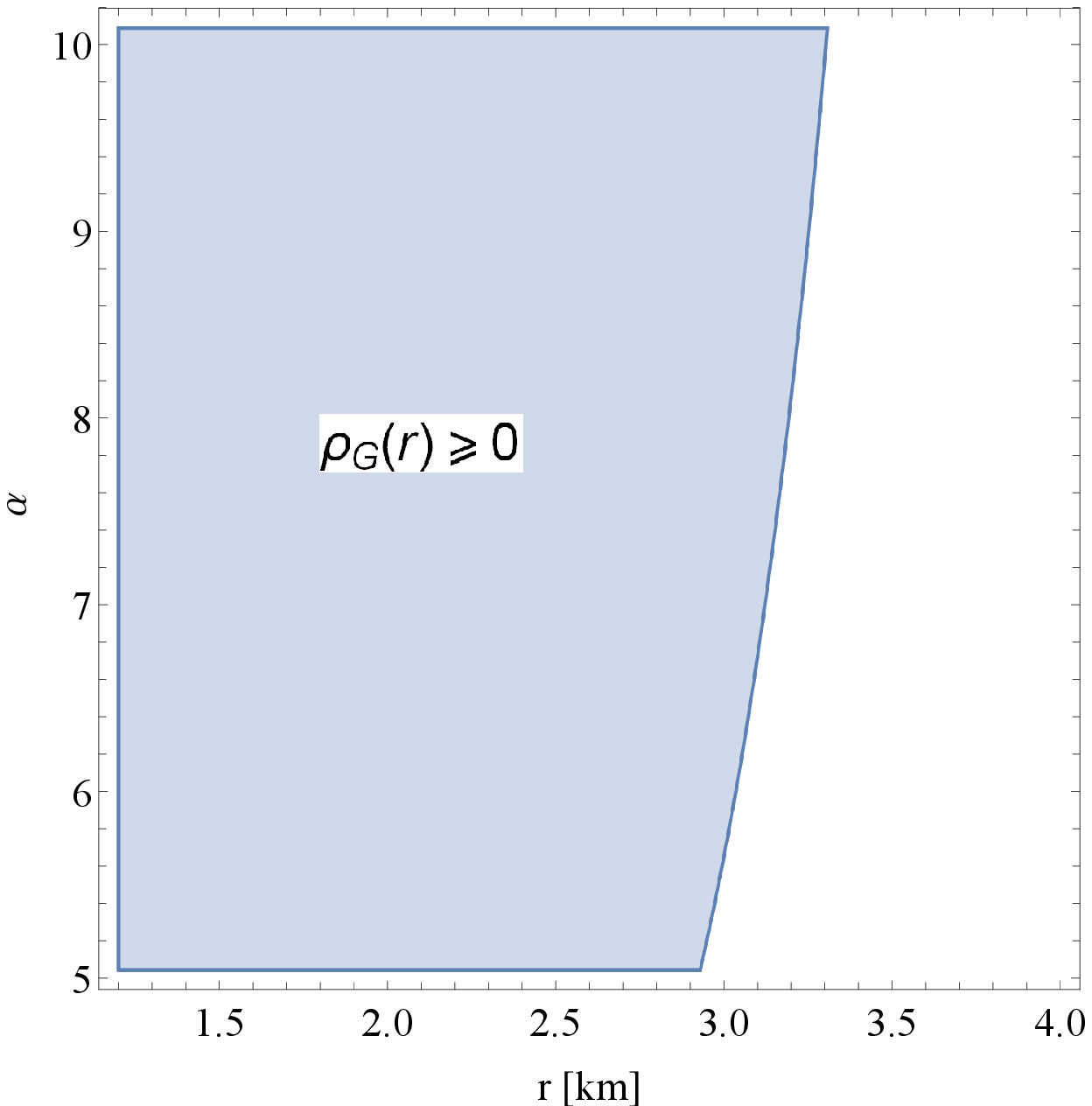}  \ 
\includegraphics[width=0.32\textwidth]{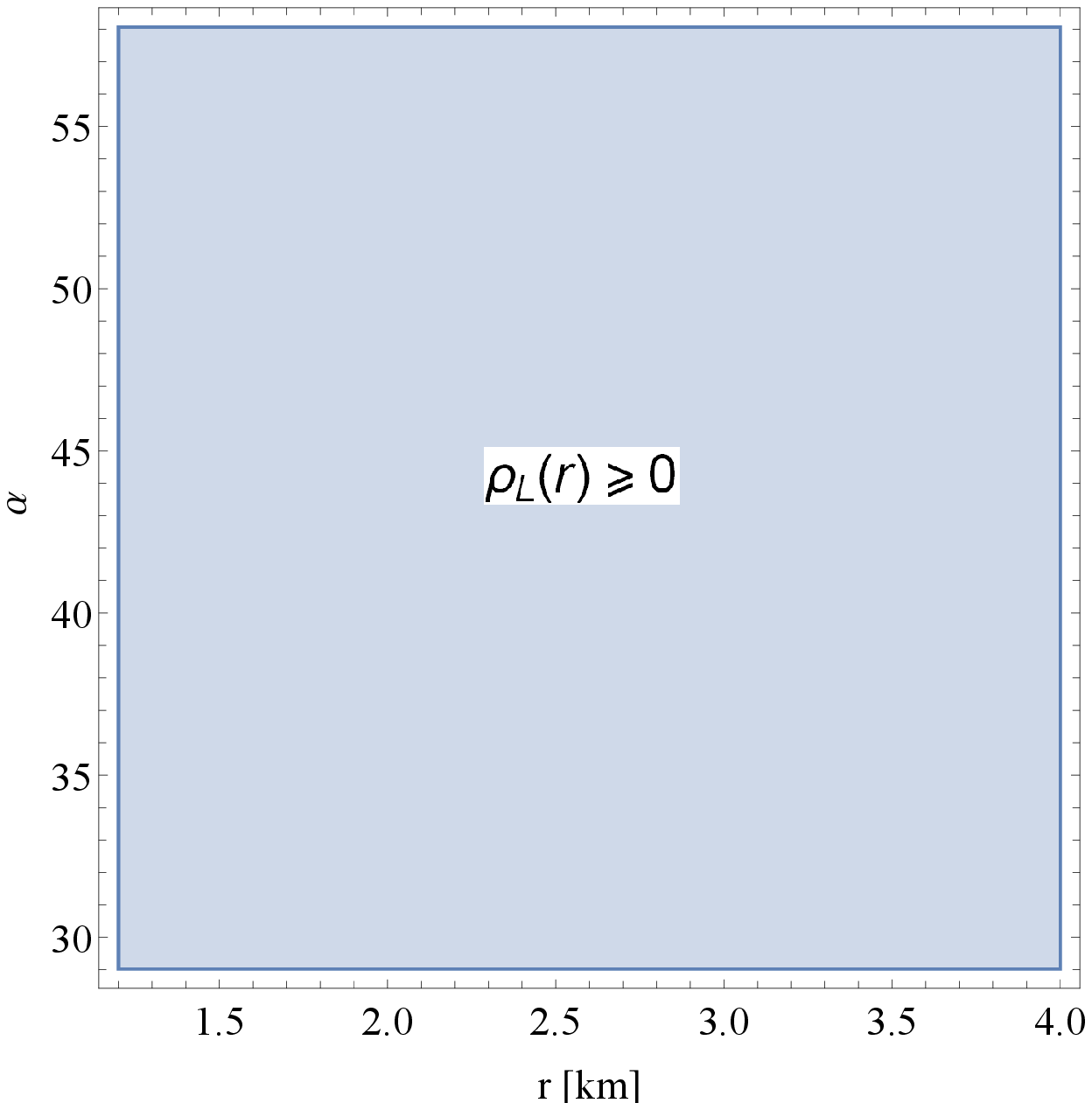}
\caption{\textbf{Upper Panels}: The NEC, WEC and SEC for the Gaussian (left) and Lorentzian (right) profiles. \textbf{Lower Panels}: The region validity for a positive Gaussian (left) and Lorentzian (right) density. These plots were constructed by considering numerical values placed in table \ref{table2}.
}\label{fig4}
\end{figure}

Let us start describing the behavior of the total density. As was pointed out, the seed density $\tilde{\rho}(r)$ is negative throughout the space--time. This is so because $\tilde{\rho}(r)$ depends on $\zeta'(r)$ and as $\zeta(r)$ is monotonically decreasing in nature, then $\zeta'(r)<0$, $\forall r\in[r_{0},+\infty)$. In considering the influence of MGD, it is possible at least to have $\rho(r)>0$ at the wormhole throat and it neighborhood. The explicit expressions for the total density for each model are given by
\begin{equation}\label{eq54}
    \rho_{G}(r)=-\frac{r^{2}_{0}}{8\pi\,r^{4}}+\frac{\alpha\mu_{G}}{\left(4\pi \mathbb{M}_{G}\right)^{3/2}}\text{Exp}\left(-\frac{r^{2}}{4\mathbb{M}_{G}}\right),
\end{equation}
and 
\begin{equation}\label{eq55}
    \rho_{L}(r)=-\frac{r^{2}_{0}}{8\pi\,r^{4}}+\frac{\alpha\sqrt{\mathbb{M}_{L}}\mu_{L}}{\pi^{2}\left(\mathbb{M}_{L}+r^{2}\right)^{2}},
\end{equation}
respectively. In the pure GR case, that is, $\alpha=0$ the magnitude of the density for each model is approximately $\rho_{G}(r_{0})=\rho_{L}(r_{0})\approx -0.0276 [\text{km}^{-2}]$. The evaluation of Eqs. (\ref{eq54}) and (\ref{eq55}) at $r_{0}$ gives $\rho(r_{0})_{G}\approx 0.0002 [\text{km}^{-2}]$ and $\rho_{L}(r_{0})\approx 0.0104 [\text{km}^{-2}]$. It is clear that the incorporation of MGD leads to a positive defined density at the wormhole throat. To further support this fact, the upper panels of Fig. \ref{fig4} (Gaussian model [left panel] and Lorentzian model [right panel]) show the trend of the total density. In contrast with the pure GR case, this quantity is positive even beyond the wormhole throat. Nevertheless, in comparing both density profiles, the Lorentzian one yields to a positive defined density everywhere not only at the mentioned regions. Otherwise, the Gaussian profile contains regions where the total density acquires a negative behavior. Lower panels in Fig. \ref{fig4} are depicting the admissible values for $\alpha$ in order to have a positive density. To understand this situation, one needs to look into the asymptotic behavior of the expressions (\ref{eq54}) and (\ref{eq55}). For the former, at larger distances from the throat, the second member at the right hand side decays faster than the first one. Therefore, the first term dominates the second one (in magnitude), causing a negative density. In the second case, although both terms in right member have the same asymptotic behavior \i.e., $\sim 1/r^{4}$ for long enough distances from the throat, the numerical coefficients of the parameters of the second term make it greater in magnitude than the first one everywhere. Thus, the total density $\rho_{L}(r)$ is always positive. This implies, that the Lorentzian profile is better than the Gaussian profile, since it is possible to obtain a positive defined total density for all $r\in[r_{0},+\infty)$.

As discussed in the flare--out condition analysis, these results are possible by constraining the magnitude and signature of $\alpha$. If one restricts the density to be positive at least at the wormhole throat, $\alpha$ is bounded from below by
\begin{equation}\label{b3}
\rho_{G}(r_{0})=-\frac{1}{8\pi r^{2}_{0}}+\frac{\alpha\mu_{G}\text{Exp}\left({-\frac{r^{2}_{0}}{4\mathbb{M}_{G}}}\right)}{8\pi^{3/2}\beta^{3/2}_{G}}\geq 0 \Rightarrow \alpha\geq \sqrt{\frac{\text{Exp}\left(\frac{r^{2}_{0}}{2\mathbb{M}_{G}}\right)\beta^{3}_{G}\pi}{r^{4}_{0}\mu^{2}_{G}}},
\end{equation}
and
\begin{equation}\label{b4}
    \rho_{L}(r_{0})=-\frac{1}{8\pi r^{2}_{0}}+\frac{\alpha\mu_{L}\sqrt{\mathbb{M}_{L}}}{\pi^{2}\left(r^{2}_{0}+\mathbb{M}_{L}\right)^{2}}\geq 0\Rightarrow \alpha\geq \frac{\pi}{8\mu_{L}}\sqrt{r^{4}_{0}\beta^{-1}_{L}+4r^{2}_{0}+6\mathbb{M}_{L}+4r^{-2}_{L}\beta^{2}_{L}+r^{-4}_{0}\beta^{3}_{L}}.
\end{equation}
As it is observed, the bounds obtained for each model are the same ones obtained when the flare--out condition is constrained (see Eqs. (\ref{b1}) and (\ref{b2})). Notwithstanding, (\ref{b3}) and (\ref{b4}) allow to take the equality whilst (\ref{b1}) and (\ref{b2}) not. Then, bounds (\ref{b1}) and (\ref{b2})) are more restrictive than (\ref{b3}) and (\ref{b4}). In any case, (\ref{b1}) and (\ref{b2})) enable to have a positive density at the throat.

Another interesting point involving the density of the wormhole structure, concern the mass of this object. As it is well-- known, the mass is obtained by a volume integration of the density over some interval. As the mass of the wormhole must be a positive, bounded and finite quantity at large enough distances. Then the density function should be at least a continuous function on the given interval to be integrable and, bounded in order to get a finite result. So, mass of the wormhole has the general form
\begin{equation}
    4\pi\int^{r}_{r_{0}}\left[\tilde{\rho}(x)+\alpha \theta^{t}_{\,t}(x)\right]x^{2}dx=\frac{1}{2}\int^{r}_{r_{0}}\left[\zeta'(x)+\alpha f'(x)\right]dx,
\end{equation}
where the above result was obtained by integrating the tt-- component of the field equations. Now, solving the right hand side one can get
\begin{equation}
\frac{1}{2}\int^{r}_{r_{0}}\left[\zeta'(x)+\alpha f'(x)\right]dx=\frac{1}{2}\left[\zeta(r)+\alpha f(r)\right]-\frac{r_{0}}{2}.
\end{equation}
Putting together previous results and following \cite{Visser:1995cc}, the mass function $m(r)$ of the wormhole is given by
\begin{equation}
    m(r)=\frac{r_{0}}{2}+4\pi\int^{r}_{r_{0}}\left[\tilde{\rho}(x)+\alpha \theta^{t}_{\,t}(x)\right]x^{2}dx=\frac{1}{2}\left[\zeta(r)+\alpha f(r)\right].
\end{equation}
Now, for each model one has the following mass function
\begin{equation} 
m_{G}(r)=\frac{r^{2}_{0}}{2r}+\alpha\mu_{G}\left[\text{Erf}\left(\frac{r}{2\sqrt{\mathbb{M}_{G}}}\right)-\left(\pi\mathbb{M}_{G}\right)^{-1/2}\text{Exp}\left(-\frac{r^{2}}{4\mathbb{M}_{G}}\right)r-\text{Erf}\left(\frac{r_{0}}{2\sqrt{\mathbb{M}_{G}}}\right)+\left(\pi\mathbb{M}_{G}\right)^{-1/2}\text{Exp}\left(-\frac{r^{2}_{0}}{4\mathbb{M}_{G}}\right)r_{0}\right],
\end{equation}
and 
\begin{equation}
m_{L}(r)= \frac{r^{2}_{0}}{2r}+\frac{2\alpha\mu_{L}}{\pi}\left[\text{arctan}\left(\frac{r}{\sqrt{\mathbb{M}_{L}}}\right)-\frac{r\sqrt{\mathbb{M}_{L}}}{\left(r^{2}+\mathbb{M}_{L}\right)}-\text{arctan}\left(\frac{r_{0}}{\sqrt{\mathbb{M}_{L}}}\right)+\frac{r_{0}\sqrt{\mathbb{M}_{L}}}{\left(r^{2}_{0}+\mathbb{M}_{L}\right)}\right]. 
\end{equation}
The total mass $M$ of the structure corresponds to the value of the mass function $m(r)$ when $r\rightarrow +\infty$. Therefore, 
\begin{equation}
    M=\lim_{r\rightarrow +\infty}m(r).
\end{equation}
For the present case one has
\begin{equation}
    M_{G}=\alpha\mu_{G}\left[1-\text{Erf}\left(\frac{r_{0}}{2\sqrt{\mathbb{M}_{G}}}\right)+\left(\pi\mathbb{M}_{G}\right)^{-1/2}\text{Exp}\left(-\frac{r^{2}_{0}}{4\mathbb{M}_{G}}\right)r_{0}\right],
\end{equation}
and
\begin{equation}
    M_{L}=\frac{2\alpha\mu_{L}}{\pi}\left[\frac{\pi}{2}-\text{arctan}\left(\frac{r_{0}}{\sqrt{\mathbb{M}_{L}}}\right)+\frac{r_{0}\sqrt{\mathbb{M}_{L}}}{\left(r^{2}_{0}+\mathbb{M}_{L}\right)}\right].
\end{equation}

Next, we shall analyze the NEC. As we are dealing with an anisotropic fluid distribution, NEC is defined by the following conditions
\begin{equation}
    \rho(r)+p_{r}(r)\geq 0 \quad \mbox{and} \quad \rho(r)+p_{\perp}(r)\geq 0.
\end{equation}
The critical point in satisfying the above conditions relies in the fulfilment of the left inequality. As demonstrated by Eq. (\ref{nec}), in the radial direction this condition is violated at the wormhole throat and its vicinity. This is so because, $p_{r}(r)<0$ and greater in magnitude than the density. In the absent of MGD ($\alpha=0$), the seed NEC in the radial direction has the following numerical value at $r_{0}$
\begin{equation}\label{necgr}
    \tilde{\rho}(r_{0})+\tilde{p}_{r}(r_{0})=-\frac{1}{4\pi r^{2}_{0}}\approx -0.0553\, [\text{km}^{-2}].
\end{equation}
On the other hand, in the transverse direction, NEC is saturated, that is
\begin{equation}
    \tilde{\rho}(r_{0})+\tilde{p}_{\perp}(r_{0})=0\, [\text{km}^{-2}].
\end{equation}
Obviously, the numerical value in radial direction of NEC, depends on the numerical value assigned to $r_{0}$. In this case, we have considered $r_{0}=1.2\,[\text{km}]$. However in the transverse direction, the numerical value is independent of numerical values assigned to any parameter. This is a general result of the chosen seed model. Despite in the tangential direction NEC is satisfied, the violation in radial direction means the violation of full NEC. Since, to satisfy the NEC, both inequalities must be simultaneously satisfied. Now, when non--commutative density profiles are incorporated via MGD, the above expressions are given by
\begin{equation}
\begin{split}
{\rho}_{G}(r)+{p}_{rG}(r)=-\frac{1}{8\pi^{3/2}r^{4}}\bigg[2r^{2}_{0}\sqrt{\pi}+2\alpha\mu_{G}\sqrt{\pi}\left(\text{Erf}\left(\frac{r}{2\sqrt{\mathbb{M}_{G}}}\right)-\text{Erf}\left(\frac{r_{0}}{2\sqrt{\mathbb{M}_{G}}}\right)r\right)+ &\\
\frac{\alpha\mu_{G}\,r}{\sqrt{\mathbb{M}_{G}}}\left(2r_{0}\text{Exp}\left(-\frac{r^{2}_{0}}{4\mathbb{M}_{G}}\right)-\frac{r\left(r^{2}+2\mathbb{M}_{G}\right)}{\beta^{2}_{G}}\text{Exp}\left(-\frac{r^{2}}{4\mathbb{M}_{G}}\right)\right)\bigg],
\end{split}
\end{equation}
\begin{equation}
\begin{split}
{\rho}_{G}(r)+{p}_{\perp G}(r)=\frac{\alpha\mu_{G}}{16\pi^{3/2}r^{3}}\bigg[2\sqrt{\pi}\left(\text{Erf}\left(\frac{r}{2\sqrt{\mathbb{M}_{G}}}\right)-\text{Erf}\left(\frac{r_{0}}{2\sqrt{\mathbb{M}_{G}}}\right)\right)+ &\\
\frac{1}{\beta^{3/2}_{G}}\left(2r_{0}\mathbb{M}_{G}\text{Exp}\left(-\frac{r^{2}_{0}}{4\mathbb{M}_{G}}\right)+r\left(r^{2}-2\mathbb{M}_{G}\right)\text{Exp}\left(-\frac{r^{2}}{4\mathbb{M}_{G}}\right)\right)\bigg],
\end{split}
\end{equation}
for the Gaussian model and, 
\begin{equation}
    \begin{split}
  {\rho}_{L}(r)+{p}_{rL}(r)=\frac{1}{4\pi^{2}r^{4}}\bigg[-r^{2}_{0}\sqrt{\pi}+2\alpha\mu_{L}\sqrt{\mathbb{M}_{L}}\left(\frac{r_{0}\sqrt{\mathbb{M}_{L}}}{r^{2}_{0}+\mathbb{M}_{L}}+\frac{\left(3r^{2}+\mathbb{M}_{L}\right)r}{\left(r^{2}+\mathbb{M}_{L}\right)^{2}}\right)r&\\+2\alpha\mu_{L}\bigg(\text{arctan}\left(\frac{r_{0}}{\sqrt{\mathbb{M}_{L}}}\right)-\text{arctan}\left(\frac{r}{\sqrt{\mathbb{M}_{L}}}\right)\bigg)r\bigg],
    \end{split}
\end{equation}
\begin{equation}
    \begin{split}
  {\rho}_{L}(r)+{p}_{\perp L}(r)=\frac{\alpha\mu_{L}}{4\pi^{2}r^{3}}\bigg[\sqrt{\mathbb{M}_{L}}\left(\frac{r_{0}}{r^{2}_{0}+\mathbb{M}_{L}}+\frac{\left(r^{2}+\mathbb{M}_{L}\right)r}{\left(r^{2}+\mathbb{M}_{L}\right)^{2}}\right)+\text{arctan}\left(\frac{r}{\sqrt{\mathbb{M}_{L}}}\right)&\\-\text{arctan}\left(\frac{r_{0}}{\sqrt{\mathbb{M}_{L}}}\right)\bigg].
    \end{split}
\end{equation}
As these expressions should be equal or greater than zero to be satisfied, as before one can restrict $\alpha$ to achieve this purpose. So, we have
\begin{eqnarray}\label{eq73}
\rho_{G}(r_{0})+p_{rG}(r_{0})&=&-\frac{1}{8\pi r^{2}_{0}}+\frac{\alpha\mu_{G}\text{Exp}\left({-\frac{r^{2}_{0}}{4\mathbb{M}_{G}}}\right)}{8\pi^{3/2}\beta^{3/2}_{G}} \geq 0\Rightarrow \alpha\geq 2\sqrt{\frac{\text{Exp}\left(\frac{r^{2}_{0}}{2\mathbb{M}_{G}}\right)\beta^{3}_{G}\pi}{r^{4}_{0}\mu^{2}_{G}}}, \quad \text{Erf}\left(\frac{r_{0}}{2\sqrt{\mathbb{M}_{G}}}\right)\in \mathbb{R}, \\ \label{eq74}
\rho_{G}(r_{0})+p_{\perp G}(r_{0})&=&\frac{\alpha\mu_{G}\text{Exp}\left({-\frac{r^{2}_{0}}{4\mathbb{M}_{G}}}\right)}{8\pi^{3/2}\beta^{3/2}_{G}}\geq0\Rightarrow \alpha\geq 0, \quad \text{Erf}\left(\frac{r_{0}}{2\sqrt{\mathbb{M}_{G}}}\right)\in \mathbb{R}.
\end{eqnarray}
As can be seen, there is a restriction on the parameter $\mathbb{M}_{G}$ in order to avoid complex values. In fact, $\mathbb{M}_{G}>0$. Now, for the Lorentzian model one gets
\begin{eqnarray}\label{eq75}
\rho_{L}(r_{0})+p_{rL}(r_{0})&=&-\frac{1}{4\pi r^{2}_{0}}+\frac{\alpha\mu_{L}\sqrt{\mathbb{M}_{L}}}{\pi^{2}\left(r^{2}_{0}+\mathbb{M}_{L}\right)^{2}}\geq 0\Rightarrow \alpha\geq \frac{\pi}{4\mu_{L}}\sqrt{r^{4}_{0}\beta^{-1}_{L}+4r^{2}_{0}+6\mathbb{M}_{L}+4r^{-2}_{L}\beta^{2}_{L}+r^{-4}_{0}\beta^{3}_{L}}, \\ \label{eq76}
\rho_{L}(r_{0})+p_{\perp L}(r_{0})&=&\frac{\alpha\mu_{L}\sqrt{\mathbb{M}_{L}}}{\pi^{2}\left(r^{2}_{0}+\mathbb{M}_{L}\right)^{2}}\geq 0\Rightarrow \alpha\geq 0.
\end{eqnarray}
For the Lorentzian model, the constant parameter
$\mathbb{M}_{L}$ also should be positive in order to avoid complex values on $\alpha$ bounds, obtained from NEC. An interesting point to be noted here, is that the bound provided by NEC in the radial direction, Eqs. (\ref{eq73}) and (\ref{eq75}), are greater in magnitude than those obtained by constraining $\alpha$ from the density (see Eqs. (\ref{b3}) and (\ref{b4})). Notwithstanding, in considering bounds (\ref{eq73}) and (\ref{eq75}), the flare--out condition would be violated, then it is not possible to have a minimal size at throat. This fact is in agreement with bounds (\ref{flare1}) and (\ref{flare2}), where it is clear that NEC and flare--out cannot be simultaneously satisfied.

Now, NEC provide the following numerical values at the throat
\begin{equation}
  {\rho}_{G}(r_{0})+{p}_{rG}(r_{0})\approx -0.0274\, [\text{km}^{-2}], \quad {\rho}_{G}(r_{0})+{p}_{\perp G}(r_{0})\approx 0,0139\, [\text{km}^{-2}],
\end{equation}

\begin{equation}
   {\rho}_{L}(r_{0})+{p}_{rL}(r_{0})\approx -0.0172\, [\text{km}^{-2}], \quad {\rho}_{L}(r_{0})+{p}_{\perp L}(r_{0})\approx 0,0190\, [\text{km}^{-2}],
\end{equation}

As can be seen, in both cases the NEC violation along the radial direction at the wormhole throat, is less in magnitude than in the pure GR case (see (\ref{necgr})). So, it can be argued that MGD helps to reduce the NEC violation, such that the wormhole is supported by a smaller amount of exotic matter than in the pure GR case.

Next, although not completely necessary, it is interesting to analyze the strong (SEC) and dominant (DEC) energy conditions, since weak energy condition (WEC) is constitutes by NEC and $\rho(r)\geq 0$ and, it was already studied. In this concern, as NEC is violated, then WEC is too. However, SEC is not violated, neither in the pure GR and MGD cases. Regarding this point, it is worth mentioning that SEC, given by 
\begin{equation}\label{sec}
    \rho(r)+p_{r}(r)\geq0, \quad    \rho(r)+p_{\perp}(r)\geq0, \quad \rho(r)+p_{r}(r)+2p_{\perp}(r)\geq 0,
\end{equation} 
is violated if NEC is transgressed. This is so because SEC implies NEC \cite{Visser:1995cc,Curiel:2014zba}. Nevertheless, the last inequality in (\ref{sec}) is always saturated, i.e., it is equal to zero. This happens only in the case when $\Phi(r)=0$ and it is a general result in GR, even when MGD scheme is present. So, at $r=r_{0}$ one gets the following bound on $\alpha$
\begin{equation}\label{eq84}
  \rho_{G}(r)+p_{rG}(r)+2p_{\perp G}(r)\geq 0  \Rightarrow \alpha \geq 0,
\end{equation}
and
\begin{equation}\label{eq85}
  \rho_{L}(r)+p_{rL}(r)+2p_{\perp L}(r)\geq 0  \Rightarrow \alpha \geq 0.
\end{equation}
The trend of NEC, WEC and SEC is depicted in Fig. \ref{fig4} (upper panels) for both models, Gaussian model (left upper panel) and Lorentzian model (right upper panel). It is clear that violation of these energy conditions is throughout the radial direction only. Since in the transverse direction they are satisfied (see green line in the upper panels of Fig. \ref{fig4}.

\begin{figure}[H]
\centering
\includegraphics[width=0.32\textwidth]{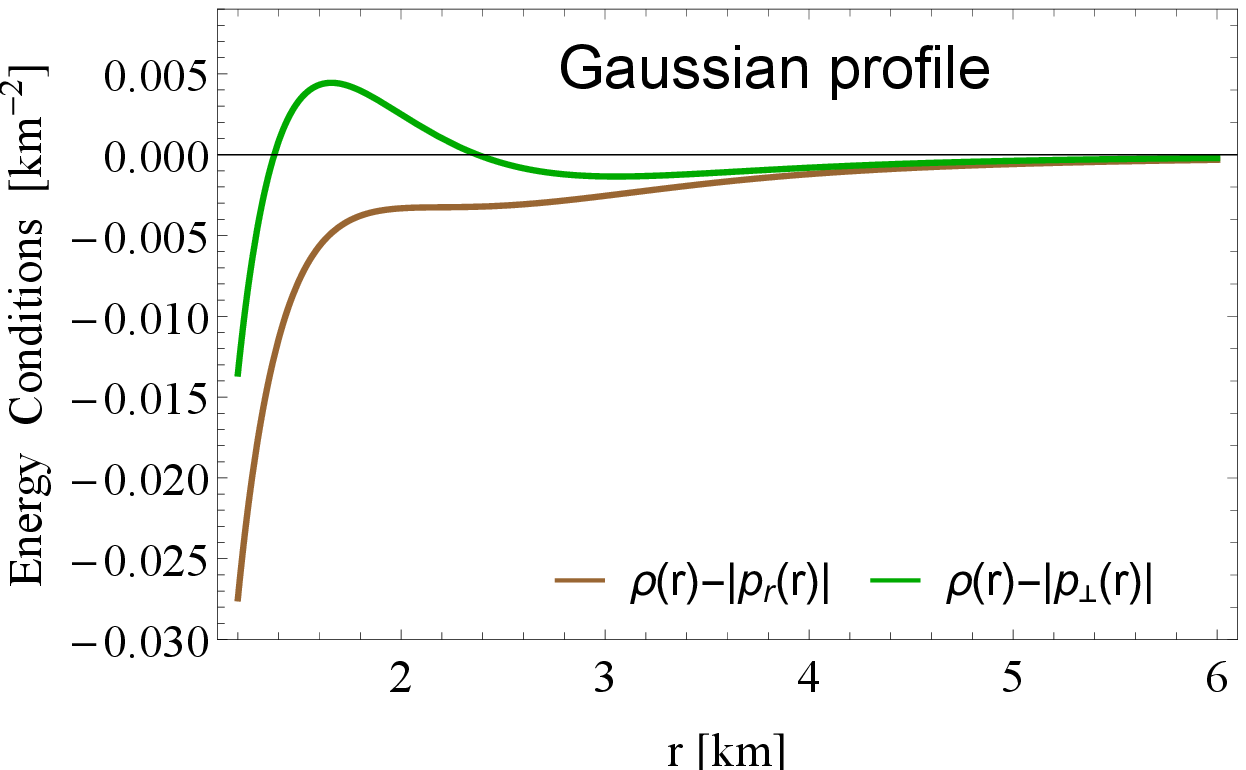}  \ 
\includegraphics[width=0.32\textwidth]{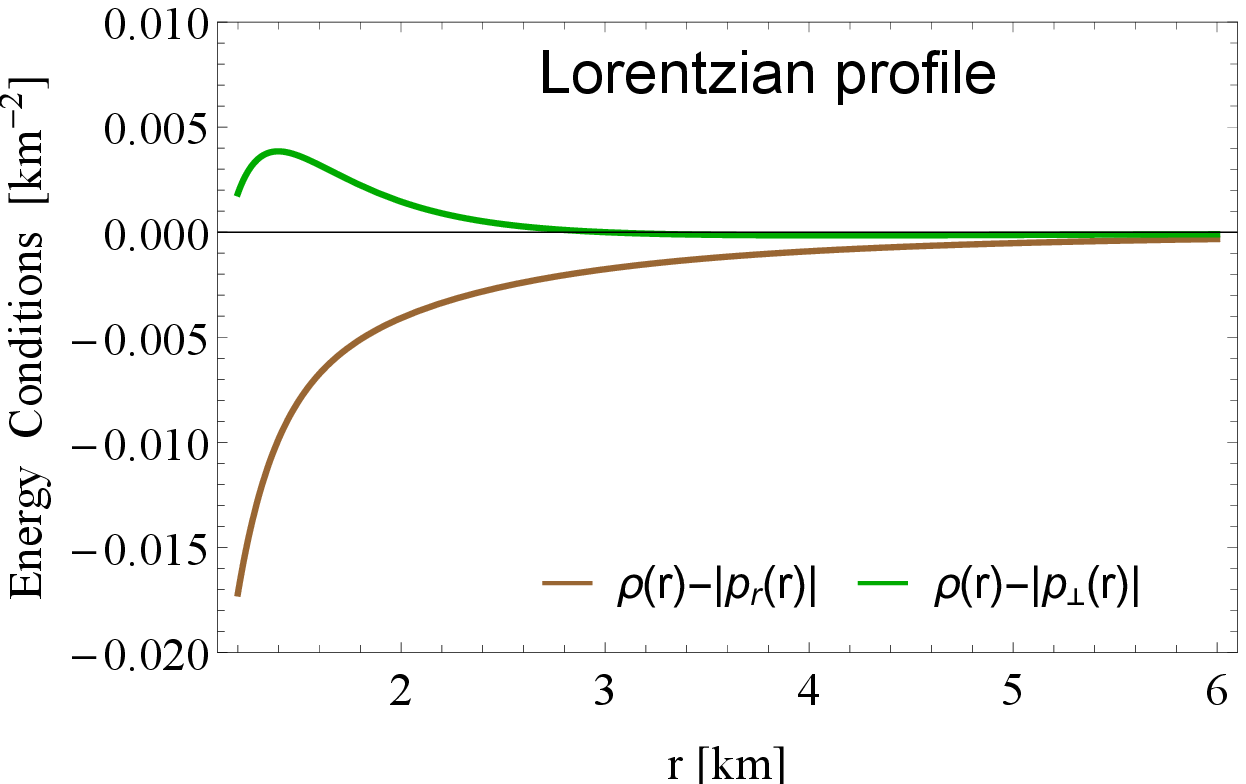}\ 
\includegraphics[width=0.3\textwidth]{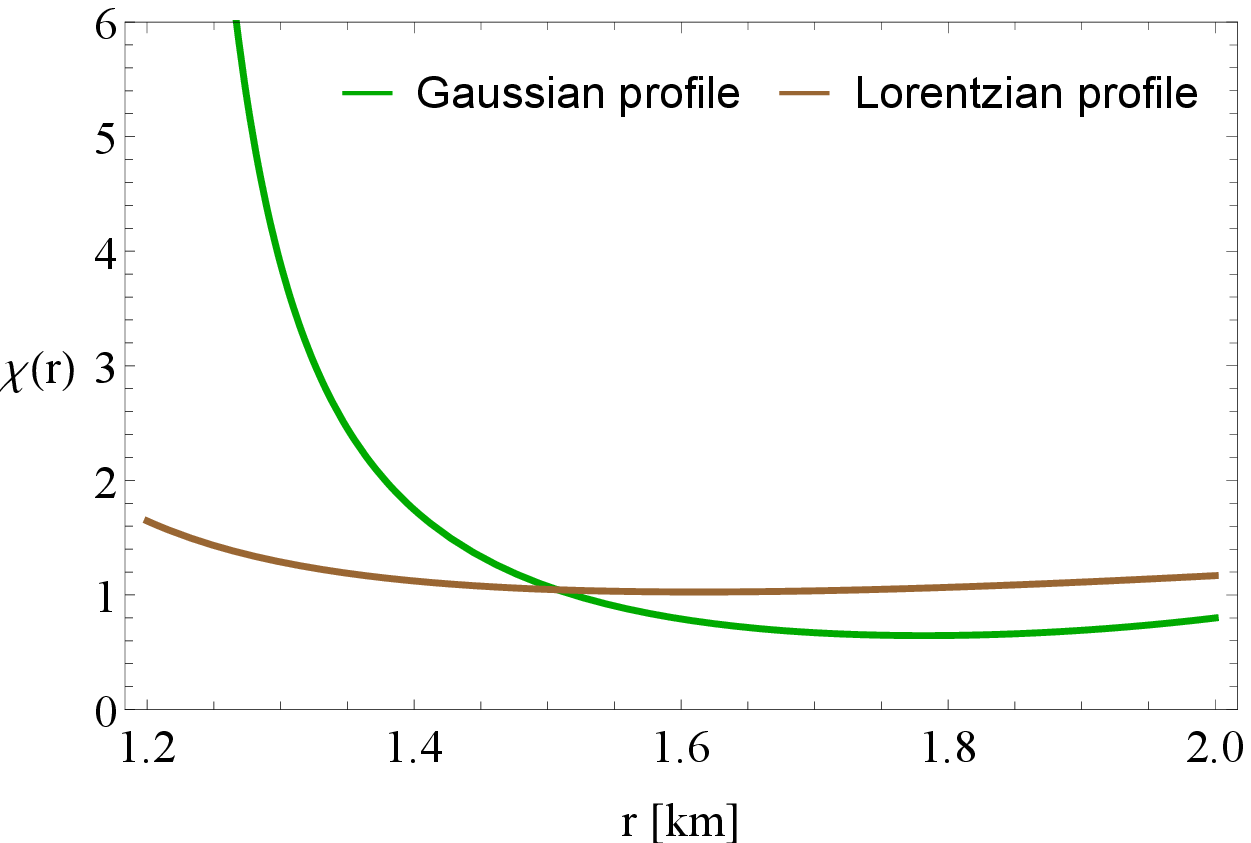}
\caption{\textbf{Left Panel}: DEC for Gaussian profile. \textbf{Middle Panel}: Dec for Lorentzian profile. \textbf{Right Panel}: The exoticity parameter $\chi(r)$ for both models. These plots were constructed by considering numerical values placed in table \ref{table2}.
}\label{fig5}
\end{figure}

On the other hand, DEC is also violated because this condition implies WEC (and thus also the NEC). Concretely, this condition reads
\begin{equation}
   \rho(r)\geq 0, \quad p_{j}(r)\in [-\rho(r),+ \rho(r)] \quad \forall j= r, \perp,  
\end{equation}
or equivalently 
\begin{equation}\label{eq83}
    \rho(r)\geq 0, \quad \rho(r)-|p_{r}(r)|\leq 0, \quad \rho(r)-|p_{\perp}(r)|\leq 0.
\end{equation}
Given that the form of DEC is to similar to the formula for WEC, this condition imposes the same bounds on $\alpha$ as WEC at least in the radial direction (see table \ref{table1} for further details).
As it is observed in Fig. \ref{fig5} (left and middle panels), DEC is violated in both directions for the Gaussian model, but only along the radial direction for the Lorentzian model. Nevertheless, from the information given in table \ref{table2}, it is clear that third inequality in (\ref{eq83}) can be satisfied by choosing a different vales for $\alpha$, for example instead of take $\alpha=18$ (Gaussian case) one can pick out $\alpha=25$.  
Finally, to further support the previous analysis, it is interesting to quantify the exotic matter distribution through a dimensional function, the so--called exoticity parameter $\chi(r)$. This parameter is defined in terms of the density and radial pressure as follows \cite{Morris1988,Morris1988A}
\begin{equation}
    \chi(r)\equiv -\frac{p_{r}(r)+
    \rho(r)}{|\rho(r)|}.
\end{equation}
Using the field equations, the dimensional parameter $\chi(r)$ can be rewritten as function of the shape function and its first derivative as follows 
\begin{equation}
    \chi(r)=\frac{b(r)-rb'(r)}{|b'(r)|r},
\end{equation}
equivalently in terms of the deformation function and flare--out condition one has
\begin{equation}\label{exo1}
\chi(r)=\frac{2\left[\zeta(r)+\alpha f(r)\right]^{2}}{r|\zeta'(r)+\alpha f'(r)|}\frac{d^{2}r}{dz^{2}}. 
\end{equation}
As flare--out condition is strictly positive, from (\ref{exo1}) it is clear that $\chi(r)>0$. Now, evaluating $\chi(r)$ at $r=r_{0}$ one arrives to
\begin{equation}
    \chi(r)\bigg{|}_{r=r_{0}}=\frac{1-\left[\zeta'(r_{0})+\alpha f'(r_{0})\right]}{|\zeta'(r_{0})+\alpha f'(r_{0})|},
\end{equation}
where the second term in the numerator is less than 1 (flare--out evaluated at the wormhole throat), hence $\chi(r_{0})>0$. The previous general results shown that if the exoticity parameter $\chi$ is positive, then the wormhole structure is supported by exotic matter. This is so because the flare--out is quite involved in the form of $\chi(r)$ having a preponderant incidence.

As it is observed in the right panel of Fig. \ref{fig5}, in both cases the exoticity parameter $\chi(r)$ takes positive values at the wormhole throat and its vicinity. This corroborates the presence of an exotic matter distribution threading the space--time. In comparing both exoticity parameter profiles, the Gaussian one overcomes the Lorentzian profile at the wormhole throat. This can be related with fact that the Lorentzian density profile, reduces in magnitude the usage of exotic matter distribution. 

\begin{table}[H]
\caption{Numerical constraints on $\alpha$ for values of $\mu_{i}$ and $\mathbb{M}_{i}$ given in TABLE \ref{table2}, provided by energy conditions and flare--out condition at the wormhole throat. }
\label{table1}
\centering
\resizebox{17.5cm}{!} {
\begin{tabular}{|c|c|c|c|c|c|} 
\cline{1-6}
Density Profile  & $\rho\geq 0$; $\rho+p_{r}\geq 0$; $\rho+p_{\perp}\geq 0$  & $\rho+p_{r}\geq 0$; $\rho+p_{\perp}\geq 0$ & $\rho+p_{r}+2p_{\perp}\geq 0$ & $\rho- |p_{r}|\geq 0$; $\rho- |p_{\perp}|\geq 0$ & $0<b'(r_{0})<1$ \\ \cline{1-6}
 Gaussian        & $\alpha \geq 17.88$; 
  $\alpha \geq 35.76$; 
  $\alpha \geq 0$ &  $\alpha \geq 35.76$; $\alpha \geq 0$   &  $\alpha \geq 0$   & $\alpha \geq 35.76$; $\alpha\geq 23.84$    & $17.88<\alpha<35.76$\\ \cline{1-6}
Lorentzian       & $\alpha \geq 29.03$; 
  $\alpha \geq 58.06$; 
  $\alpha \geq 0$ &  $\alpha \geq 58.06$; $\alpha \geq 0$   &  $\alpha \geq 0$   & $\alpha \geq 58.06$; $\alpha\geq 38.71$    & $29.03<\alpha<58.06$        \\ \cline{1-6}
\end{tabular}}
\end{table}

\section{Conclusions}\label{sec4}

{In this article, new wormhole solutions in the GR scenario have been provided. To obtain these solutions, the seminal Morris--Thorne wormhole space--time is deformed by employing gravitational decoupling, by means of the so--called minimal geometric deformation approach. To close the $\theta$--sector, we have interpreted the temporal component $\theta^{t}_{\ t}(r)$ as the noncommutative density profile, specifically the Gaussian (\ref{Gauss}) and Lorentzian profiles (\ref{Lorentz}). This information in conjunction with the equation (\ref{eq16}) allow to obtain the decoupler function for each model, (\ref{f1}) for the Gaussian profile and (\ref{f2}) for the Lorentzian one. As the throat condition $b(r_{0})=r_{0}$ must be respected, the deformation function should be zero at the wormhole throat $r_{0}$, given rise to the minimally deformed shape functions (\ref{eq34}) and (\ref{eq35}).  }

{Interestingly, the resulting space--times are preserving the asymptotically flat behavior of the seed solution. This fact can be corroborated in the Fig. \ref{fig1}, where the inverse of the radial metric potential \i.e., $1-b(r)/r$ tends to the unity when $r\rightarrow +\infty$. Furthermore, the embedding diagrams in Fig. \ref{fig3} also are corroborating the mentioned behavior. Actually, the asymptotically flat behavior of the salient solution not need in general to be preserved, since it depends on how the $\theta$--sector is closed. On the other hand, it should be noted that this features of the seed model can be maintained if from the very beginning the deformation function $f(r)$ is carefully selected with the following asymptotic behavior: $f(r)\sim \mathcal{O}\left(1/r\right)$ subject to the condition $f(r_{0})=0$. On the other hand, from Fig. \ref{fig6} it is clear the effects of $\alpha$--terms coming form the $\theta$--sector, have on the geometry of the space--time. In this case as shown by the 3--dimensional embedding diagram, $\alpha$--terms are modifying the proper length $l(r)$ of the wormhole throat.  }

{Another important geometrical feature in studying traversable wormholes, is the so--called flare--out condition. In this case, for both solutions this condition is satisfied. What is more, to assure the validity of the flare--out, we have restricted at the wormhole throat the $\alpha$ parameter using the condition $b'(r_{0})<1$ (for each model see equations (\ref{flare1}) and (\ref{flare2})). As can be observed in Fig. \ref{fig1}, the behavior of the flare--out condition at the wormhole throat and throughout the space--time is satisfied.    }

{Concerning the matter distribution and the satisfaction of energy conditions. It is remarkable to note that, in the GR arena, it is not possible to avoid the violation of energy conditions. Nevertheless, constraining the $\alpha$ parameter, it is possible to obtain a positive defined density (see constraints (\ref{b3}) and (\ref{b4}) for each model). This can be seen in the upper panels of Fig. \ref{fig4}, while in the lower panels it is displayed the region validity of a positive density along the full space--time. On the other hand, from energy conditions, NEC, WEC, SEC and DEC, the coupling $\alpha$ also has been constrained, in order to see how to overcome the usage of exotic matter. In this regard, constraints (\ref{eq73}), (\ref{eq74}) and (\ref{eq84}) for Gaussian profile and (\ref{eq75}), (\ref{eq76}) and (\ref{eq85}) for Lorentzian profile, are providing opposite constraints on $\alpha$ (specially those given by NEC and WEC ), in contrast with the flare--out condition (\ref{flare1}) and (\ref{flare2}). This is so because, energy conditions and flare--out cannot be satisfied at the same footing in the GR context, if one wants to obtain traversable wormhole solutions. In other words, exotic matter is an essential ingredient in the construction of traversable wormholes in the GR scenario.   }

{Notwithstanding, by controlling the magnitude of parameter $\alpha$, the usage of exotic matter can be greatly reduced. This fact is observed in Figs. \ref{fig1} (upper panels) and \ref{fig5} (left and middle panels), where energy conditions are taking small negative values at the wormhole throat and its neighborhood in comparison with what is occurring in GR. To further support the presence of exotic matter, the so--called exoticity parameter has been displayed in the right panel of Fig. \ref{fig5}. In this concern, it is clear that the Lorentzian profile offers a better scenario in comparison with the Gaussian one, since the exoticity parameter in the Gaussian profile case overcomes the Lorentzian one, taking higher positive values. As it is well--known, a positive exoticity parameter means exotic matter presence in the wormhole space--time and its magnitude reflects or quantifies in some way, how much exotic matter there is in the structure.}

{As final comment, one can say that gravitational decoupling by MGD, is a good tool to obtain new an interesting wormhole solutions, either in GR or in other gravity theories. However, it should be pointed out that more analysis is necessary, to further support the feasibility of these models. For example, it is necessary to perform and stability analysis. This problem will be faced elsewhere.}

\section*{Acknowledgement} 
F. Tello-Ortiz thanks the financial support by projects ANT--2156 and SEM 18--02 at the Universidad de Antofagasta. BM acknowledges IUCAA, Pune, India for providing support through the Visiting Associateship Program.

\bibliography{references}

\end{document}